\documentclass[11pt,a4paper]{article}
\pdfoutput=1
\usepackage{jcappub}

\def\wisk#1{\ifmmode{#1}\else{$#1$}\fi}

\def\lsim   {\wisk{_<\atop^{\sim}}}

\def\deg    {\wisk{^\circ}}
\def\ddeg   {\wisk{{\rlap.}^\circ}}

\title{The Primordial Inflation Explorer (PIXIE):
A Nulling Polarimeter for Cosmic Microwave Background Observations} 

\author[1]{A. Kogut}
\author[1,2]{D.J. Fixsen}
\author[1]{D.T. Chuss}
\author[3]{J. Dotson}
\author[1]{E. Dwek}
\author[4]{M. Halpern}
\author[4]{G.F. Hinshaw}
\author[5]{S.M. Meyer}
\author[1]{S.H. Moseley}
\author[6]{M.D. Seiffert}
\author[7]{D.N. Spergel}
\author[1]{and E.J. Wollack}

\affiliation[1]{Code 665, 			
		Goddard Space Flight Center,	
		Greenbelt, MD 20771 USA }
\affiliation[2]{University of Maryland,		
		College Park, MD 20742 USA }
\affiliation[3]{MS 245-6,			
		Ames Research Center,		
		Mofett Field, CS 94035 USA }
\affiliation[4]{University of British Columbia,	
		Vancouver, BC V67 1Z1 Canada }
\affiliation[5]{University of Chicago,		
		5801 South Ellis Ave,		
		Chicago, IL 60637 USA }
\affiliation[6]{M/S 169-506,			
		Jet Propulsion Laboratory,	
		Pasadena, CA 91109 USA }

\emailAdd{Alan.J.Kogut@nasa.gov}

\abstract{
The Primordial Inflation Explorer (PIXIE)
is an Explorer-class mission 
to measure the gravity-wave signature of primordial inflation 
through its distinctive imprint on the linear polarization 
of the cosmic microwave background.  
The instrument consists of a
polarizing Michelson interferometer
configured as a nulling polarimeter
to measure the difference spectrum
between orthogonal linear polarizations
from two co-aligned beams.
Either input can view the sky
or a temperature-controlled
absolute reference blackbody calibrator.
PIXIE will map the absolute intensity and linear polarization
(Stokes $I$, $Q$, and $U$ parameters)
over the full sky
in 400 spectral channels spanning 2.5 decades in frequency 
from 30 GHz to 6 THz (1 cm to 50 $\mu$m wavelength).  
Multi-moded optics provide background-limited sensitivity
using only 4 detectors,
while the highly symmetric design
and multiple signal modulations
provide robust rejection of potential systematic errors.
The principal science goal is the detection and characterization 
of linear polarization from an inflationary epoch in the early universe, 
with tensor-to-scalar ratio $r < 10^{-3}$ at 5 standard deviations.  
The rich PIXIE data set will also constrain physical processes 
ranging from Big Bang cosmology 
to the nature of the first stars 
to physical conditions within the interstellar medium of the Galaxy.
}

\keywords{CMBR experiments,
CMBR polarisation,
inflation,
reionization}

\notoc

\begin{document}
\maketitle

% Begin main text
\flushbottom

\section{Introduction}
A central principle in modern cosmology
is the concept of inflation,
which posits a period of exponential expansion
in the early universe shortly after the Big Bang.
The many $e$-foldings of the scale size during inflation
force the geometry of space-time to asymptotic flatness
while dilating quantum fluctuations in the inflaton potential
to the macroscopic scales responsible for seeding 
large-scale structure in the universe.
Inflation provides a simple, elegant solution
to multiple problems in cosmology,
but it relies on extrapolation of physics
to energies greatly exceeding direct experimentation
in particle accelerators.

The polarization of the cosmic microwave background (CMB)
provides a direct test of inflationary physics.
CMB polarization results from Thomson scattering of CMB photons
by free electrons.
A quadrupolar anisotropy in the radiation incident on each electron
creates a net polarization in the scattered radiation.
There are only two possible sources for such a quadrupole:
either an intrinsic temperature anisotropy 
or the differential redshift caused by a gravity wave
propagating through an isotropic medium.
The two cases can be distinguished by their different spatial signatures.
Temperature perturbations are scalar quantities;
their polarization signal must therefore be curl-free.
Gravity waves, however, are tensor perturbations
whose polarization includes both gradient and curl components.
In analogy to electromagnetism,
the scalar and curl components are
often called ``E'' and ``B'' modes.
Only gravity waves induce a
curl component: 
detection of a B-mode signal in the CMB polarization field
is recognized as a ``smoking gun'' signature of inflation,
testing physics at energies inaccessible through any other means
\citep{rubakov/etal:1982,
fabbri/pollock:1983,
abbott/wise:1984,
polnarev:1985,
davis/etal:1992,
grishchuk:1993,
kamionkowski/etal:1997,
seljak/zaldarriaga:1997}.

\begin{figure}[t]
\centerline{
\includegraphics[width=5.0in]{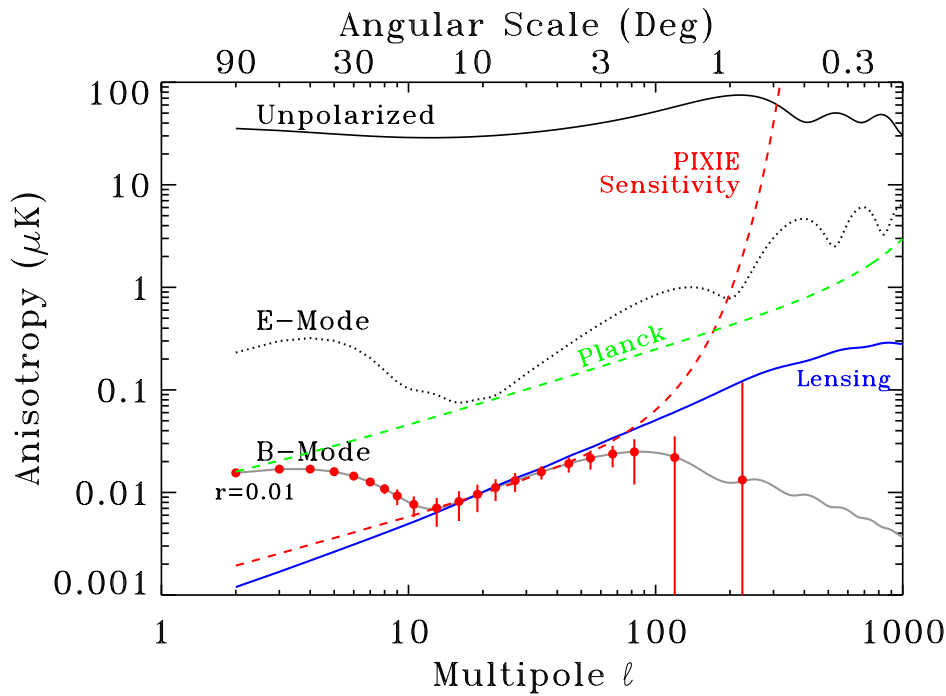}}
\caption{
Angular power spectra for 
unpolarized,
E-mode,
and B-mode polarization
in the cosmic microwave background.
The dashed red line shows the PIXIE sensitivity
to B-mode polarization 
at each multipole moment $\ell \sim 180\deg/\theta$.
The sensitivity estimate assumes a 4-year mission
and includes the effects of foreground subtraction
within the cleanest 75\% of the sky
combining PIXIE data at frequencies $\nu < 600$ GHz.
Red points and error bars show the response 
within broader $\ell$ bins
to a B-mode power spectrum with amplitude $r = 0.01$.
PIXIE will reach the confusion noise (blue curve)
from the gravitational lensing of the E-mode signal
by cosmic shear along each line of sight,
and has the sensitivity and angular response
to measure even the minimum predicted B-mode power spectrum
at high statistical confidence.
}
\label{power_spectra_fig}
\end{figure}

Figure \ref{power_spectra_fig}
shows the amplitude of CMB polarization
as a function of angular scale.
At the degree angular scales
characteristic of the horizon at decoupling,
the unpolarized temperature 
anisotropy is typically 80 $\mu$K.
These fluctuations in turn
generate E-mode polarization,
which at amplitude $\sim$3 $\mu$K 
is only a few percent of the temperature fluctuations.
The B-mode amplitude from gravity waves is unknown.
This amplitude is related directly to the energy scale $V_*$ of inflation,
$V_* = r(0.003~M_{pl})^4$
\citep{turner/white:1996},
where
$M_{pl} = 1.22 \times 10^{19}$ GeV is the Planck mass
and $r = T/S$ is the power ratio of gravity waves to scalar fluctuations.
If inflation results from Grand Unified Theory physics
(energy $\sim ~10^{16}$ GeV),
the B-mode amplitude should be in the range 1 to 100 nK.
Recent WMAP results suggest likely values 30-100 nK,
toward the upper range of GUT inflation
\citep{komatsu/etal:2009}.
Signals at this amplitude
could be detected by  a dedicated polarimeter,
providing a direct, model-independent measurement 
of the energy scale of inflation.

%--------------------------------------------------------------------------
% Figure 1: Power spectra of temperature and polarization anisotropy
%--------------------------------------------------------------------------
%--------------------------------------------------------------------------

Detecting the gravity-wave signature
in polarization will be difficult.
As recognized in multiple reports
\citep{bock/etal:2006,
dunkley/etal:2009,
dodelson/etal:2009},
there are three fundamental challenges:

\begin{itemize}

\item 
{\bf Sensitivity~}
The gravity-wave signal is faint
compared to the fundamental sensitivity limit
imposed by photon arrival statistics.
Even noiseless detectors suffer from this photon-counting limit;
the only solution is to collect more photons.

\item 
{\bf Foregrounds~}
The gravity-wave signal is faint compared to 
the polarized Galactic synchrotron and dust foregrounds.
Separating CMB from foreground emission
based on their different frequency spectra requires multiple frequency channels.

\item 
{\bf Systematic Errors~}
The gravity-wave signal is faint compared to both
the unpolarized CMB anisotropy and the dominant E-mode polarization.
Accurate measurement of the B-mode polarization
requires strict control of instrumental effects
that could alias these brighter signals into a false B-mode detection.

\end{itemize}

\noindent
Satisfying the simultaneous requirements of sensitivity, 
foreground discrimination, and immunity to systematic errors
presents a technological challenge. 
In this paper, we describe an instrument 
capable of measuring the CMB and diffuse Galactic foregrounds 
with background-limited sensitivity
in over 400 frequency channels using only 4 detectors. 

\section{Instrument Description}
The Primordial Inflation Explorer (PIXIE)
is an Explorer-class mission to detect and characterize 
the polarization signal from an inflationary epoch
in the early Universe.
PIXIE combines multi-moded optics with a 
Fourier Transform Spectrometer
to provide breakthrough sensitivity for CMB polarimetry
using only four semiconductor detectors.
The design addresses each of the principal challenges 
for CMB polarimetry.
A multi-moded ``light bucket'' provides nK sensitivity
using only four detectors. 
A polarizing Fourier Transform Spectrometer (FTS)
synthesizes 400 channels across 2.5 decades in frequency 
to provide unparalleled separation of CMB from Galactic foregrounds. 
PIXIE's highly symmetric design
enables operation as a nulling polarimeter to provide
the necessary control of instrumental effects.

%--------------------------------------------------------------------------
% Figure 2: PIXIE concept
%--------------------------------------------------------------------------
\begin{figure}[t]
\centerline{
\includegraphics[height=5.0in]{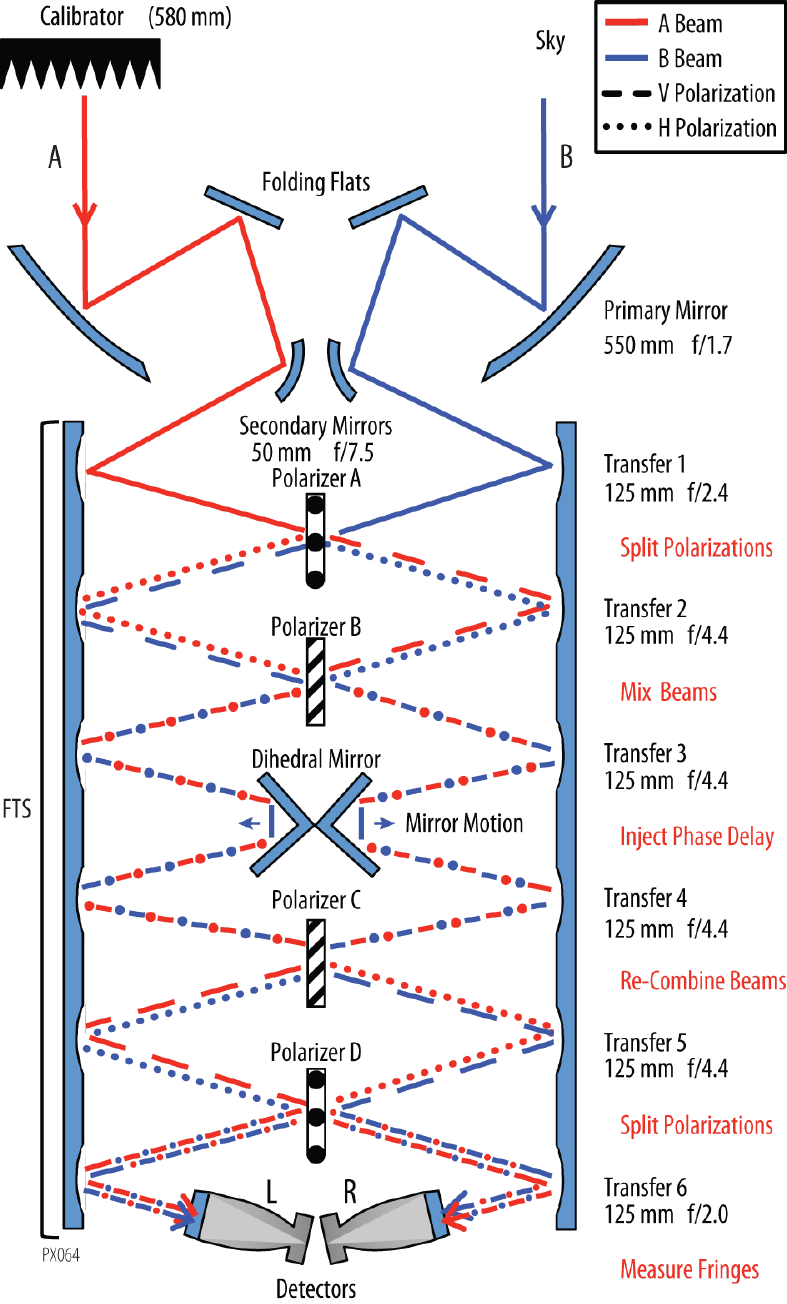}}
\caption[PIXIE Concept]
{PIXIE optical signal path.
As the dihedral mirrors move,
the detectors measure a fringe pattern proportional to the
Fourier transform of the difference spectrum
between orthogonal polarization states from the two input beams
(Stokes Q in instrument coordinates).
A full-aperture blackbody calibrator
can move to block either input beam,
or be stowed to allow both beams to view the same patch of sky.
}
\label{pixie_concept_fig}
\end{figure}
%--------------------------------------------------------------------------

Figure \ref{pixie_concept_fig} shows the instrument concept.
Two off-axis primary mirrors 550 mm in diameter
produce twin beams co-aligned with the spacecraft spin axis.
A folding flat and 50 mm secondary mirror
route the beams to the FTS.
A set of six transfer mirror pairs, 
each imaging the previous mirror to the following one, 
shuttles the radiation through a series of polarizing wire grids.
Polarizer A transmits vertical polarization 
and reflects horizontal polarization, 
separating each beam into orthogonal polarization states. 
A second polarizer (B) with wires oriented 45\deg ~relative to grid A 
mixes the polarization states.
A Mirror Transport Mechanism (MTM) moves back-to-back dihedral mirrors 
to inject an optical phase delay. 
The phase-delayed beams re-combine (interfere) at Polarizer C. 
Polarizer D (oriented the same as A) 
splits the beams again and routes them to
two multi-moded concentrator feed horns. 
Each concentrator is square to preserve linear polarization
and contains a pair of identical bolometers,
each sensitive to a single linear polarization 
but mounted at 90\deg ~to each other 
to measure orthogonal polarization states. 
To control stray light,
all internal surfaces except the active optical elements
are coated with a microwave absorber
\citep{wollack/etal:2008},
forming a blackbody cavity isothermal with the sky.

Each of the four detectors measures an interference fringe pattern
between orthogonal linear polarizations from the two input beams.
Let $\vec{E} = E_x \hat{x} + E_y \hat{y}$ 
represent the electric field incident from the sky.
The power at the detectors
as a function of the mirror position $z$
may be written
\begin{eqnarray}
P_{Lx} &=& \frac{1}{2} ~\int(E_{Ax}^2+E_{By}^2)+(E_{Ax}^2-E_{By}^2) \cos(4z\omega /c)d\omega    \nonumber \\
P_{Ly} &=& \frac{1}{2} ~\int(E_{Ay}^2+E_{Bx}^2)+(E_{Ay}^2-E_{Bx}^2) \cos(4z\omega /c)d\omega	\nonumber \\
P_{Rx} &=& \frac{1}{2} ~\int(E_{Ay}^2+E_{Bx}^2)+(E_{Bx}^2-E_{Ay}^2) \cos(4z\omega /c)d\omega    \nonumber \\
P_{Ry} &=& \frac{1}{2} ~\int(E_{Ax}^2+E_{By}^2)+(E_{By}^2-E_{Ax}^2) \cos(4z\omega /c)d\omega~,
\label{full_p_eq}
\end{eqnarray}
(Appendix A),
where
$\omega$ is the angular frequency of incident radiation,
L and R refer to the detectors in the left and right concentrators,
and A and B refer to the two input beams
(Fig \ref{pixie_concept_fig}).

% -------------- Table 1: Optical Parameters --------------
\begin{table}[t]
{
\small
\caption{Optical Parameters}
\label{optic_param}
\begin{center}
\begin{tabular}{| l | c | l | }
\hline 
Parameter			&	Value	& 	Notes \\
\hline 
Primary Mirror Diameter		& 55 cm			& Sets beam size on sky	\\
Etendu				& 4 cm$^2$ sr		& 2.7 times larger than FIRAS \\
Beam Diameter			& 2\ddeg6~tophat	& Equivalent 1\ddeg6~Gaussian FWHM \\
Throughput			& 82\%			& Excludes detector absorption \\
Detector Absorption		& 54\%			& Reflective backshort	\\
\hline
Mirror Stroke			& $\pm$2.6 mm peak-peak	& Phase delay $\pm$10 mm	\\
Spectral Resolution		& 15 GHz		& Set by longest mirror stroke \\
Highest Effective Frequency	& 6 THz			& Spacing in polarizing grids \\
\hline 
Detector NEP			& $0.7 \times 10^{-16}$ W Hz$^{-1}$	&	\\
System NEP			& $2.7 \times 10^{-16}$ W Hz$^{-1}$	& Background limit \\
\hline
\end{tabular}
\end{center}
}
\end{table}
%------------------------------------------------------------

The term modulated by the mirror scan is proportional to the
Fourier transform of the frequency spectrum
for Stokes $Q$ linear polarization
in instrument-fixed coordinates.
Rotation of the instrument about the beam axis
interchanges $\hat{x}$ and $\hat{y}$ 
on the detectors.  The sky signal 
(after the Fourier transform)
then becomes
\begin{eqnarray}
S(\nu)_{Lx} &=& \frac{1}{4} 
	\left[ ~I(\nu)_A - I(\nu)_B 
	+ Q(\nu)_{\rm sky} \cos 2\gamma + U(\nu)_{\rm sky} \sin 2\gamma ~ \right] 
	\nonumber \\
S(\nu)_{Ly} &=& \frac{1}{4}
	\left[ ~I(\nu)_A - I(\nu)_B 
	- Q(\nu)_{\rm sky} \cos 2\gamma - U(\nu)_{\rm sky} \sin 2\gamma ~\right]
	\nonumber \\
S(\nu)_{Rx} &=& \frac{1}{4} 
	\left[ ~I(\nu)_B - I(\nu)_A 
	+ Q(\nu)_{\rm sky} \cos 2\gamma + U(\nu)_{\rm sky} \sin 2\gamma ~ \right] 
	\nonumber \\
S(\nu)_{Ly} &=& \frac{1}{4}
	\left[ ~I(\nu)_B - I(\nu)_A 
	- Q(\nu)_{\rm sky} \cos 2\gamma - U(\nu)_{\rm sky} \sin 2\gamma ~\right]~,
\label{diff_spectra_eq}
\end{eqnarray}
where
$I = \langle E_x^2 + E_y^2 \rangle$,
$Q = \langle E_x^2 - E_y^2 \rangle$, 
and
$U = 2 {\rm Re} \langle E_x E_y \rangle$
are the Stokes polarization parameters,
$\gamma$ is the spin angle,
and $S(\nu)$ denotes the synthesized frequency spectrum
with bins $\nu$ set by the fringe sampling.

Table \ref{optic_param} summarizes the instrument optics.
Cryogenic pupil stops at the primary mirror
and field stops at the transfer mirrors
limit the etendu to 4 cm$^2$ sr
to produce a circular tophat beam with diameter 2\ddeg6\footnote{
The angular smoothing from the tophat beam 
(window function in $\ell$)
may be approximated
by a 1\ddeg6 ~Gaussian full width at half maximum.}.
Throughput from the entrance aperture 
to the detector is 82\%,
evenly split among geometric loss,
diffractive loss,
and absorption on the wire grids.
A low-pass filter on each folding flat blocks zodiacal light.
The optics, including the square concentrator,
preserve polarization:
orthogonal polarization states at the detector
remain orthogonal within 3\deg
~when projected to the sky.

%--------------------------------------------------------------------------
% Figure 3: Thermal cartoon
%--------------------------------------------------------------------------
\begin{figure}[t]
\centerline{
\includegraphics[width=4.0in]{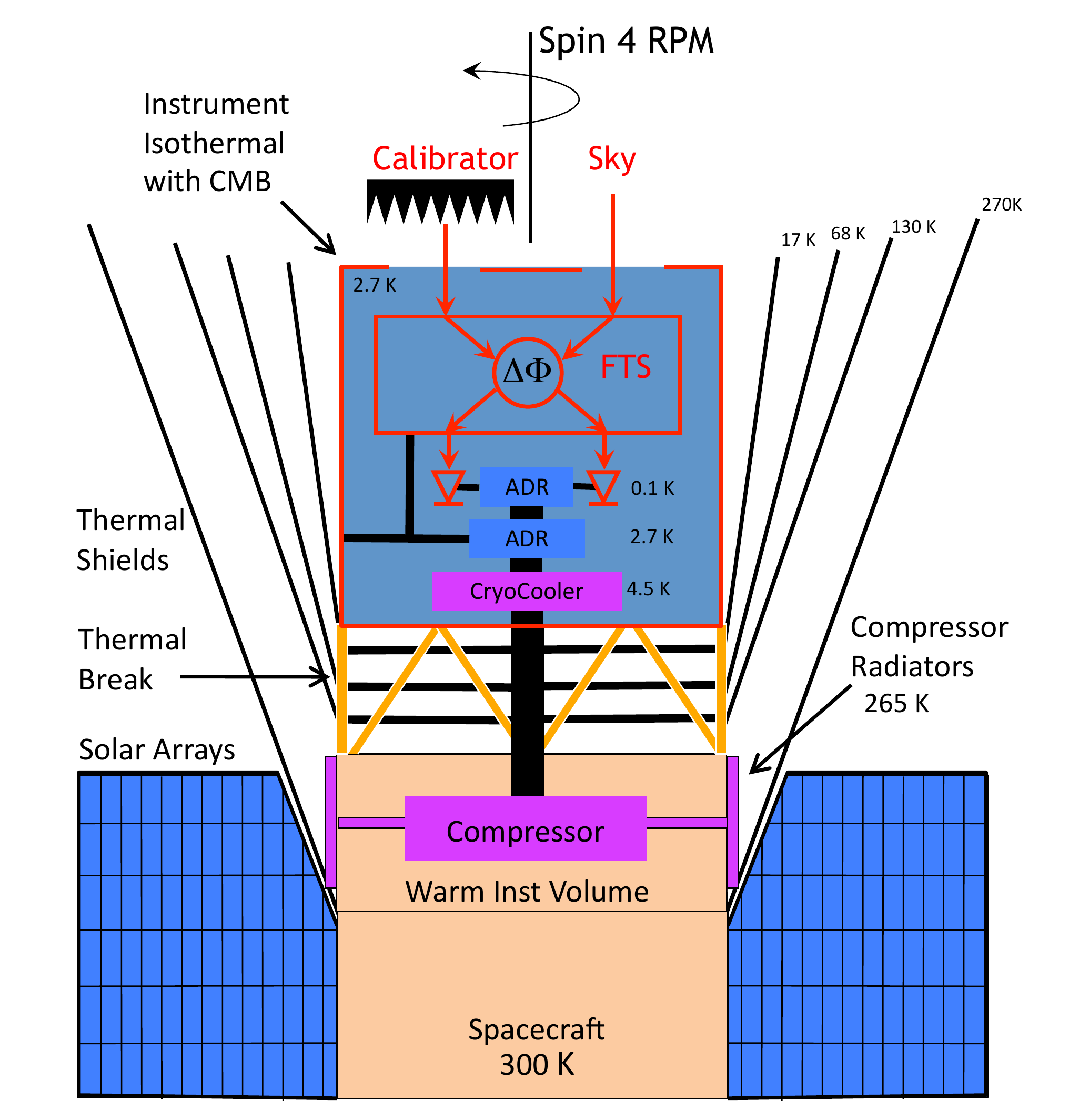}}
\caption{
Cryogenic layout for the PIXIE instrument.
An ADR and mechanical cryo-cooler
maintain the instrument and enclosure at 2.725 K,
isothermal with the CMB.
A set of concentric shields surrounds the instrument
to prevent heating by the Sun or Earth.
}
\label{cryo_fig}
\end{figure}
%--------------------------------------------------------------------------

PIXIE operates as a nulling polarimeter:
when both beams view the sky,
the instrument nulls all unpolarized emission
so that the fringe pattern responds only to the sky polarization.
The resulting null operation greatly reduces
sensitivity to systematic errors
from unpolarized sources.
Normally the instrument collects light
from both co-aligned telescopes.
A full-aperture blackbody calibrator
can move to block either beam,
replacing the sky signal in that beam
with an absolute reference source,
or be stowed to allow both beams to view the same sky patch.
The calibrator temperature is maintained near 2.725 K
and is changed $\pm 5$ mK every other orbit
to provide small departures from null
as an absolute reference signal.
When the calibrator blocks either beam,
the fringe pattern encodes information
on both the temperature distribution on the sky (Stokes I)
as well as the linear polarization.
Interleaving observations with and without the calibrator
allows straightforward transfer of the absolute calibration scale
to linear polarization,
while providing a valuable cross-check of the polarization solutions
obtained in each mode.

The PIXIE design differs radically from kilo-pixel focal plane arrays,
but shares a number of similarities with 
the Far Infrared Absolute Spectrophotometer (FIRAS) instrument
on NASA's Cosmic Background Explorer (COBE) mission
\citep{mather:1982a,mather/etal:1990,mather/etal:1993}.
Both instruments use a polarizing Michelson interferometer
with free-standing wire grid polarizers
to measure the frequency spectrum over several decades.
PIXIE unfolds the optics so that each photon interacts
with each grid and mirror once instead of twice,
thus requiring four grids to FIRAS' two.
The larger PIXIE etendu
(4 cm$^2$ sr compared to 1.5 cm$^2$ sr for FIRAS)
and lower bolometer NEP
($0.7 \times 10^{-16} ~{\rm W~Hz}^{-1/2}$
compared to
$2 \times 10^{-15} ~{\rm W~Hz}^{-1/2}$ for FIRAS)
provide a factor of 76 in improved sensitivity.
FIRAS compared a single sky beam to an internal blackbody calibrator,
and occasionally inserted an external calibrator for absolute calibration.
The PIXIE optical path is fully symmetric,
with two sky beams incident on the FTS.
The PIXIE external calibrator can be moved to block either beam
or stowed so that both beams view the sky.
Dichroic splitters divided the FIRAS output 
into a high-frequency and low-frequency band,
using a total of four identical detectors
(left high, left low, right highn and right low).
PIXIE divides each ouput by polarization,
also utilizing four identical detectors
(left $\hat{x}$,
left $\hat{y}$,
right $\hat{x}$, and
right $\hat{y}$).
Each PIXIE detector measures the difference
between orthogonal linear polarizations
from opposite sides of the instrument.

%--------------------------------------------------------------------------
% Figure 4: Observatory and mission concept
%--------------------------------------------------------------------------
\begin{figure}[t]
\centerline{
\includegraphics[width=6.5in]{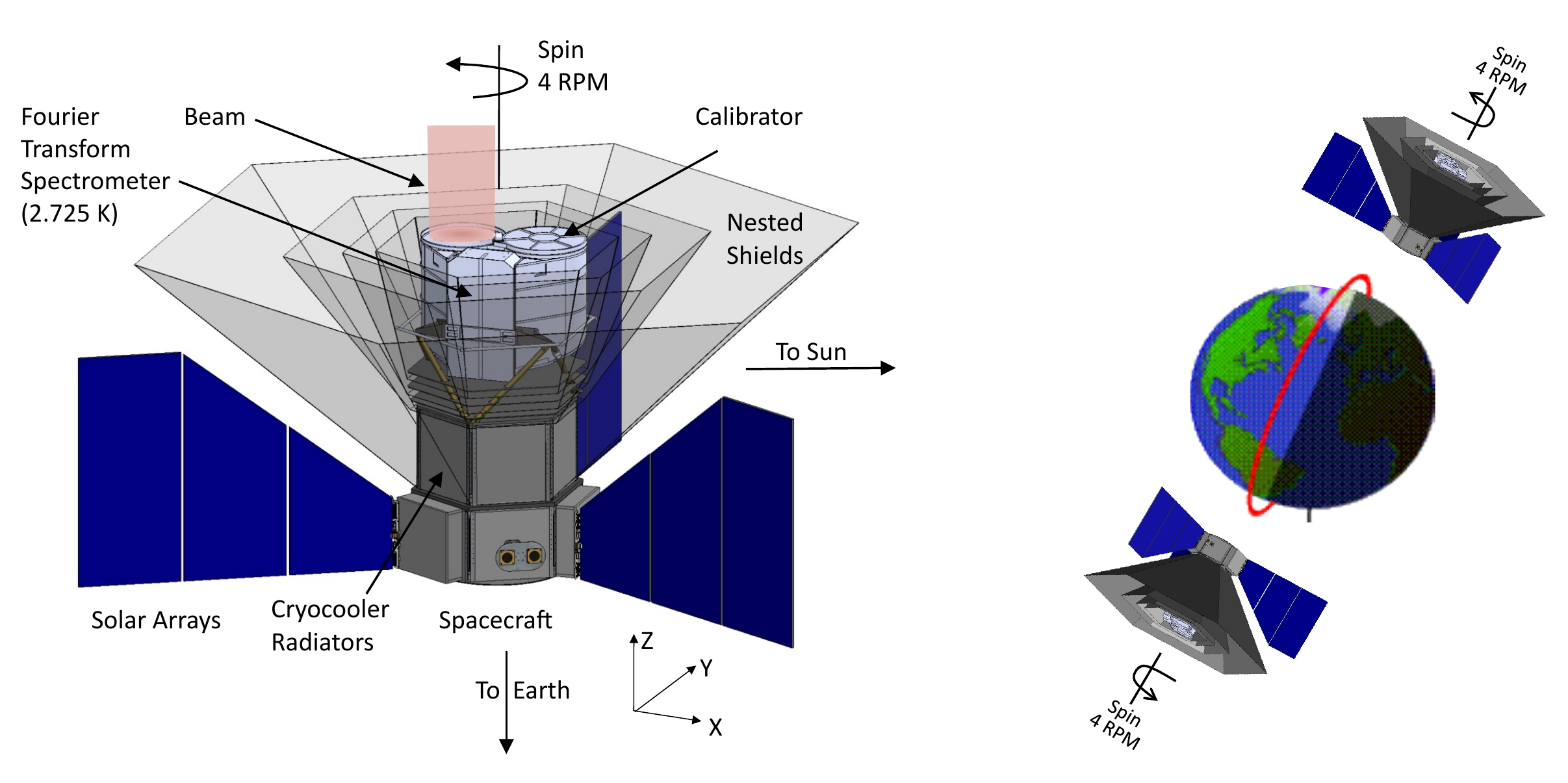}}
\caption{
PIXIE observatory and mission concept.
The instrument is maintained at 2.725 K
and is surrounded by shields to block radiation from the Sun or Earth.
It observes from a 660 km polar sun-synchronous terminator orbit.
The rapid spin and interferometer stroke
efficiently separate Stokes I, Q, and U parameters
independently within each pixel
to provide a nearly diagonal covariance matrix.
}
\label{orbit_fig}
\end{figure}
%--------------------------------------------------------------------------

The frequency multiplex advantage inherent in the FTS spectrometer 
also means that each synthesized frequency bin 
contains noise defined 
not by the background intensity evaluated at that frequency,
but by the integrated background intensity
incident on the detector. 
For a ground-based experiment,
the resulting noise from atmospheric emission 
would be unacceptable compared to other techniques. 
PIXIE is optimized for a space environment
where the nulling FTS polarimeter
provides significant advantages
for both instrument simplicity 
and control of systematic errors.

Figure \ref{cryo_fig}
shows the thermal schematic of the instrument.
An adiabatic demagnetization refrigerator (ADR)
maintains the telescope and FTS at 2.725 K,
in thermal equilibrium with the CMB.
A second ADR cools the detectors and associated feed horn assembly
to 0.1 K.
An absorbing enclosure,
also maintained at 2.725 K,
surrounds the optics to
control stray light,
so that radiation leaving the detector (in a time-reversed sense)
terminates either on the sky
or on a blackbody enclosure
with spectrum nearly identical to the sky.
The ADR rejects heat to a mechanical cryo-cooler at 4.5 K,
which in turn rejects the heat to a dedicated radiator
viewing deep space.
A hexapod thermal break structure
provides thermal isolation between the cryogenic 
and warm portions of the instrument.
A set of concentric shields surrounds the instrument
and cryocooler radiator
to prevent heating by the Sun or Earth
while also providing passive cooling
to the thermal break structure.

Figure \ref{orbit_fig}
shows the observatory and mission concept.
PIXIE will launch into a 660 km polar sun-synchronous orbit
with 6 AM or 6 PM ascending node
to provide full-sun operation.
The instrument spins at 4 RPM with the spin axis maintained
91\deg ~from the Sun line
and as close as possible to the zenith 
consistent with the Solar pointing requirement.
The instrument thus observes a great circle each orbit,
while the orbit precession of 1\deg ~per day
achieves full sky coverage in each 6-month observing period.
The detector sampling, mirror stroke, and spacecraft spin
are fast compared to the orbital motion of the beam
across the sky,
eliminating the need for
pixel-to-pixel differences
in the data analysis.

\begin{figure}[t]
\centerline{
\includegraphics[height=4.0in]{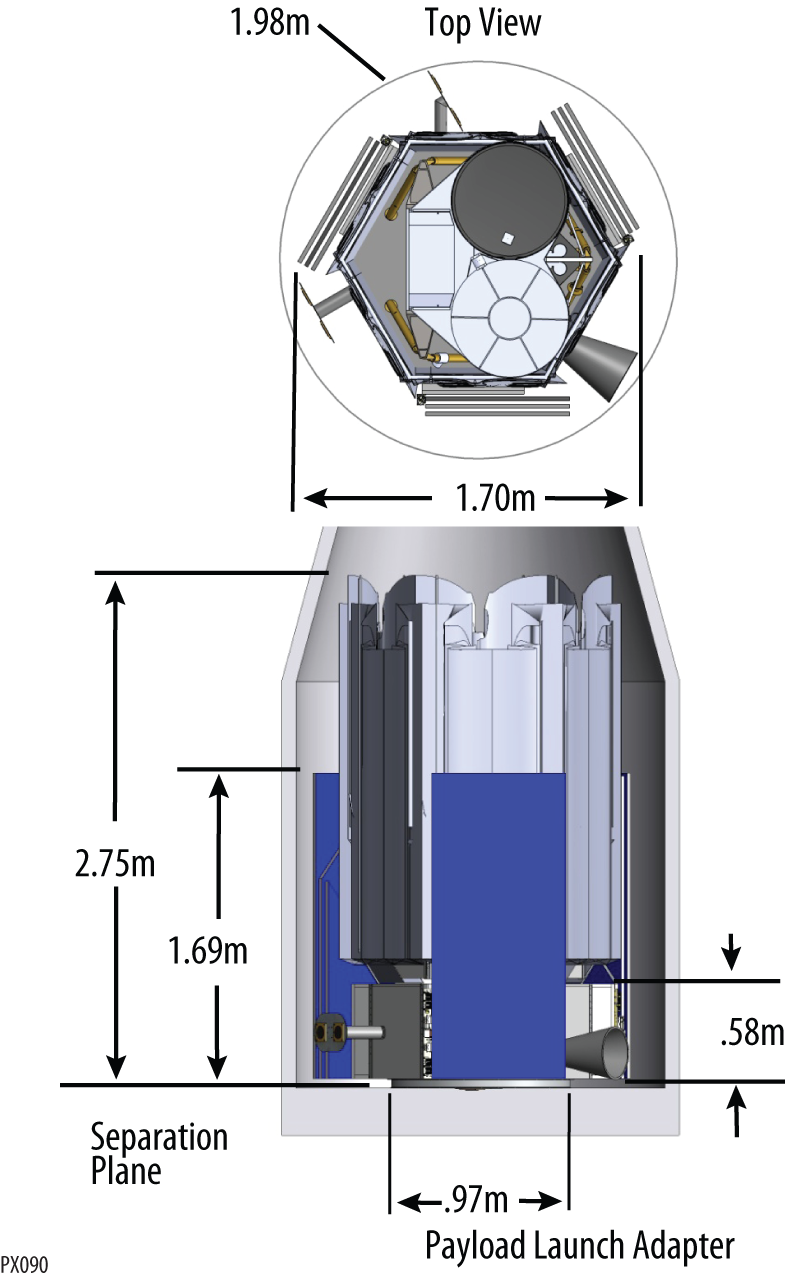}}
\caption{
The PIXIE observatory fits easily
within the Explorer 92" fairing.
}
\label{launch_fig}
\end{figure}

PIXIE is small mission,
well within the capabilities of NASA's Explorer program.
Figure \ref{launch_fig}
shows the observatory in launch configuration.
The observatory size is set by the two 55 cm primary mirrors,
and fits easily within the 92" fairing
of either the Taurus or Athena launch vehicles.
PIXIE is technologically mature
and could launch as early as 2017.

%--------------------------------------------------------------------------
% Figure 5: Observatory in launch configuration
%--------------------------------------------------------------------------
%--------------------------------------------------------------------------

\section{Instrument Performance}

\subsection{Sensitivity}

The gravity-wave signal is faint
compared to the fundamental sensitivity limit
imposed by photon arrival statistics.
Currently fielded instruments
search for the inflationary signature
using kilo-pixel arrays of transition-edge superconducting bolometers
\citep{filippini/etal:2010,
reichborn/etal:2010,
chuss/etal:2010}.
Strictly speaking, however,
the need is for more {\it photons},
not necessarily more {\it detectors}.
The noise equivalent power (NEP) of photon noise
in a single linear polarization is given by
\begin{equation}
{\rm NEP}^2_{\rm photon} = {2A\Omega \over c^2} {(kT)^5\over h^3}
	\int \alpha \epsilon f
	  ~\frac{x^4}{e^x-1} 
	  ~\left( 1 + \frac{\alpha \epsilon f}{e^x-1} \right) ~dx ,
\label{mather_12a}
\end{equation}
where
$x=h\nu/kT$,
$\nu$ is the observing frequency,
$A$ is the detector area,
$\Omega$ is the detector solid angle,
$\alpha$ is detector absorptivity,
$T$ is the physical temperature of the source,
$\epsilon$ is the emissivity of the source,
and
$f$ is the power transmission through the optics
\citep{mather:1982b}.
For a fixed integration time $\tau$ the detected noise is simply
\begin{equation}
\delta P = \frac{ {\rm NEP}}{ \sqrt{\tau / 2}}
\label{noise_eq}
\end{equation}
where the factor of 2 accounts for the conversion
between the frequency and time domains.
The noise at the detector
may in turn
be referred to the specific intensity on the sky,
\begin{equation}
\delta I_\nu = \frac{ \delta P }
		      { A\Omega ~\Delta \nu ~(\alpha \epsilon f) }
\label{i_noise}
\end{equation}
where 
$\Delta \nu$ is the observing bandwidth.

The light-gathering ability of an instrument is specified by its etendu 
$A \Omega$.
Two points are worth noting. 
Increasing the etendu for a single detector
increases the photon noise,
${\rm NEP}~\propto~(A \Omega)^{1/2}$,
thereby decreasing the relative contribution of the intrinsic
detector (phonon) noise.
But since the signal increases linearly with etendu,
the signal-to-noise ratio {\it improves} as $(A \Omega)^{1/2}$.
Increasing the etendu relaxes detector noise requirements
while simultaneously improving
the overall system sensitivity to sky signals.

The multi-moded PIXIE optics provide sensitivity
comparable to kilo-pixel focal plane arrays
while requiring only 4 semiconductor bolometers.
For diffraction-limited single-mode optics, 
the etendu and wavelength are related as 
$A \Omega = \lambda^2$ 
so that the beam size scales with the observing wavelength. 
For multi-moded optics, however, 
the beam size is fixed and the number of modes $N$ 
scales as 
$N = A \Omega / \lambda^2$.
Multi moded optics thus allow a considerable increase in sensitivity
compared to single-moded designs of comparable size.
The improvement is large enough
to allow precision measurement of 
the gravity-wave signature in polarization
using a handful of detectors.
Over just the frequency range 30--600 GHz where the CMB is brightest, 
each PIXIE detector measures 22,000 independent modes of the electric field.

The fringe pattern measured at each detector
samples the Fourier transform of the frequency spectrum
of the difference between one linear polarization from the A-side beam
and the orthogonal linear polarization from the B-side beam
(Eq. \ref{full_p_eq}).
The frequency bins in the synthesized spectra
are set by the mirror throw and detector sampling.
As the mirror moves,
we obtain $N_s$ detector samples
over an optical path length $\pm \Delta L$.
The Fourier transform of the sampled fringe pattern
returns frequencies
$n ~ \times c/(2 \Delta L)$
where
$n = 0, 1, 2, ..., N_s/2$.
The path length (optical stroke)
thus determines the width of the frequency bins
in the synthesized spectra,
while the number of detector samples within each optical stroke
determines the number of frequency bins
and thus the highest sampled frequency.
With $N_s = 1024$
and $\Delta L = 1$ cm,
we obtain 512 synthesized frequency bins
of width 15 GHz each.
The corresponding physical mirror movement
$\Delta z = \Delta L / [4 \cos(\alpha) \cos(\delta/2)]$ = $\pm$2.58 mm
accounts for the folded optics
as well as the off-axis optical path
$(\alpha = 15\deg$)
and beam divergence
($\delta = 6.5\deg$).

Table \ref{optic_param} summarizes the instrument optical parameters.
The instrument NEP of 
$2.7 \times 10^{-16}~{\rm W~Hz}^{-1/2}$
is dominated by photon noise
with only minor contribution
from the intrinsic detector noise.
With the calibrator deployed over either aperture,
the instrument measures both polarized and unpolarized emission
(Stokes $I$, $Q$, and $U$).
With both beams open to the sky,
the instrument is insensitive to unpolarized emission
but has twice the sensitivity to polarized signals\footnote{
Replacing the blackbody calibrator emission with sky emission
in one beam
leaves the noise nearly unchanged
but doubles the sky signal incident on the FTS.}.
Reducing the mirror stroke 
when the calibrator is stowed
further improves sensitivity
(the wider frequency bins do not degrade 
foreground subtraction
since no polarized line emission is anticipated).
Including these effects
and averaging over the four detectors,
the combined instrument sensitivity to either unpolarized or polarized emission
within each synthesized frequency bin is
\begin{eqnarray}
\delta I_\nu^I &=& 2.4 \times 10^{-22} 
~{\rm W ~m}^{-2} ~{\rm sr}^{-1} ~{\rm Hz}^{-1}	\nonumber \\
\delta I_\nu^{QU} &=& 3.4 \times 10^{-22} 
~{\rm W ~m}^{-2} ~{\rm sr}^{-1} ~{\rm Hz}^{-1}
\label{spectral_noise_i_mode}
\end{eqnarray}
for a one-second integration
with the calibrator deployed over either aperture,
and
\begin{eqnarray}
\delta I_\nu^{QU} &=& 0.5 \times 10^{-22} 
~{\rm W ~m}^{-2} ~{\rm sr}^{-1} ~{\rm Hz}^{-1}
\label{spectral_noise_q_mode}
\end{eqnarray}
when the calibrator is stowed
(Appendix B).
PIXIE will spend approximately 30\% of the observing time
with the calibrator deployed and 60\% with the calibrator stowed.
The remaining 10\% includes both high-temperature calibration
of the Galactic dust signal
and lost observing time.

% -------------- Table 2: Observatory NET and NEQ --------------
\begin{table}[t]
\caption{Observatory CMB Sensitivity}
\label{neq_table}
\begin{center}
\begin{tabular}{| l | c | c |}
\hline 
Observing Mode	&  NET			&	NEQ \\
		& ($\mu{\rm K}~s^{1/2}$) &  ($\mu{\rm K}~s^{1/2}$) \\
\hline
Calibrator Deployed	& 13.6		& 19.2 \\
Calibrator Stowed	& ---		& 5.6 \\
\hline
\end{tabular}
\end{center}
\end{table}
%------------------------------------------------------------

\begin{figure}[t]
\centerline{
\includegraphics[width=4.5in]{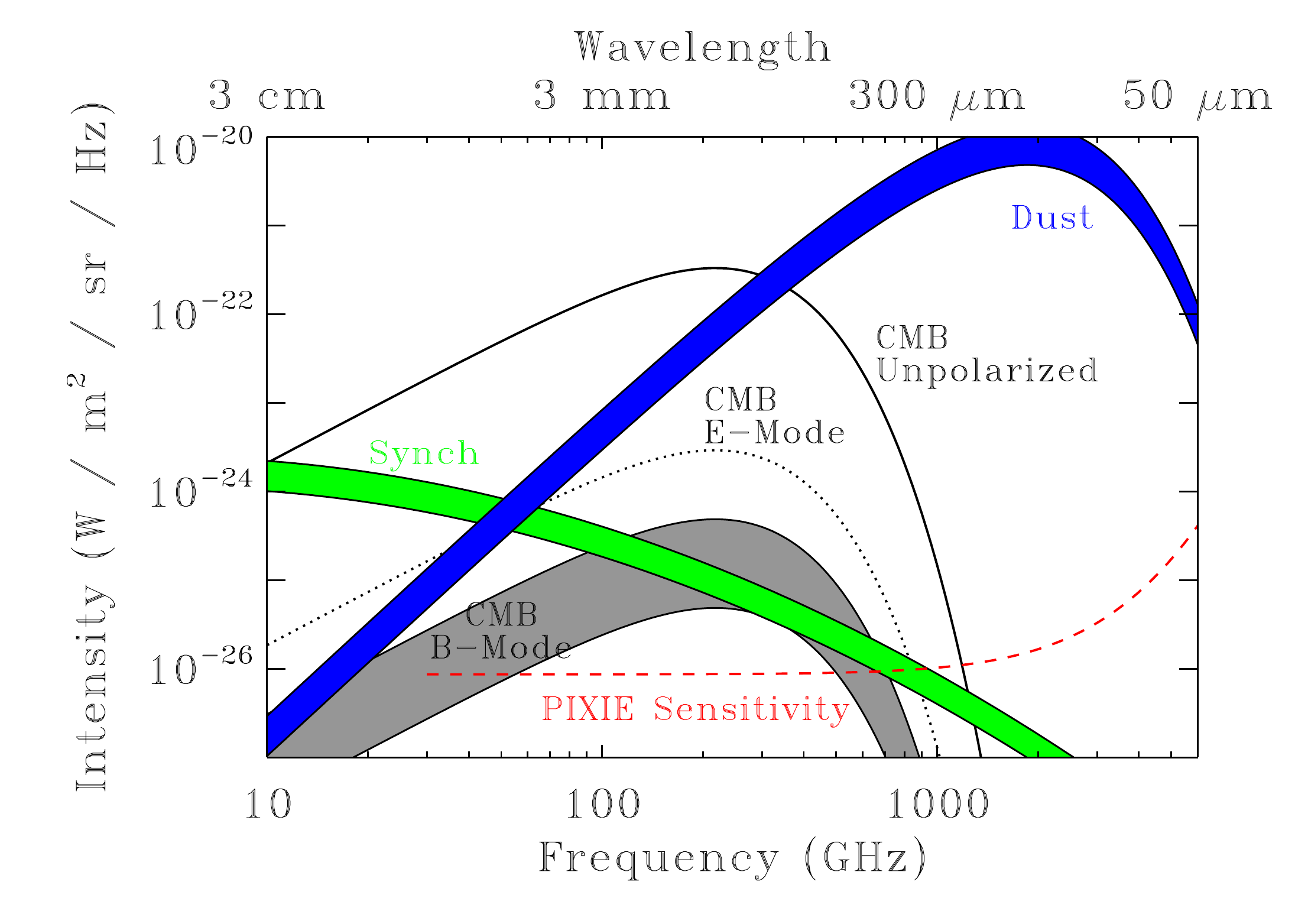}}
\caption[CMB and Foreground Polarization]
{RMS anisotropy for the CMB and polarized foregrounds.
The grey band shows the range of predicted amplitudes
for the primordial gravity-wave signal.
PIXIE combines multi-moded optics
with a Fourier Transform Spectrometer
to achieve high sensitivity in 400 spectral channels
spanning 2.5 decades in frequency.
}
\label{foreground_fig}
\end{figure}

Eqs \ref{spectral_noise_i_mode} and \ref{spectral_noise_q_mode}
give the specific intensity within each synthesized frequency bin.
The wire grid polarizers become inefficient in reflection
at wavelengths $\lambda < 60~\mu$m
defined by the $30~\mu$m wire spacing,
limiting the effective frequency coverage
to 400 bins 
from 30~GHz to 6~THz
(Fig \ref{foreground_fig}).
For continuum sources like the CMB
we may integrate over multiple bins to further improve sensitivity.
Table \ref{neq_table}
shows the resulting sensitivity to a CMB source.
A 4-year mission achieves rms sensitivity
70 nK within each $1\deg \times 1\deg$~pixel.

Cosmological foregrounds present a natural sensitivity limit.
Primordial density perturbations
source E-mode polarization.
Gravitational lensing from the mass distribution
along each line of sight creates a shear 
that distorts the orientation of the primordial polarization field, 
analogous to the shear in galaxy orientations 
observed along the line of sight toward distant massive clusters. 
By perturbing the polarization orientation, 
gravitational lensing mixes the E and B modes 
and transforms a small fraction of the dominant E-mode polarization 
into a B-mode.
On angular scales $\theta > 0\ddeg5$, 
the lensed B-mode signal is well approximated by a random white noise field 
on the sky
(Fig \ref{power_spectra_fig}).
This sky noise is present along every line of sight 
and adds in quadrature with the instrument noise.
PIXIE will reach reach this cosmological ``noise floor''
beyond which little additional gain can be realized.

The PIXIE sensitivity allows robust detection and characterization
of the inflationary B-mode signal.
Averaged over the cleanest 75\% of the sky,
PIXIE will detect the gravity-wave signal 
$r < 10^{-3}$ at 5 standard deviations,
two orders of magnitude more sensitive than Planck
and a factor of ten below the predicted signal
for GUT-scale inflation.
The resulting sensitivity allows characterization
of the B-mode angular power spectrum
in $\sim$20 bins at  multipoles
$\ell < 100$
limited by the 2\ddeg6~diameter beam
(Figure~\ref{power_spectra_fig}).

%--------------------------------------------------------------------------
% Figure 6: Foreground polarization
%--------------------------------------------------------------------------
%--------------------------------------------------------------------------

\subsection{Foreground Subtraction}
The most demanding challenge to detecting cosmological B modes
is likely to be confusion from foreground signals.
PIXIE has both the sensitivity and multi-frequency lever arm
necessary to map foreground emission independently
in each sky pixel.

On the large angular scales of interest to PIXIE,
foregrounds are dominated by polarized emission
from the Milky Way's interstellar medium
\citep{dunkley/etal:2009,
fraisse/etal:2008}.
Polarized emission within the Galaxy is dominated by two main sources.
Synchrotron radiation from cosmic ray electrons
accelerated in the Galactic magnetic field
approximates a power-law spectrum
\begin{equation}
I(\nu)_{\rm synch} \propto \nu^{\beta_s}
\label{synch_eq}
\end{equation}
with $\beta_s \sim -0.7$.
Thermal emission from dust grains
follows a greybody spectrum,
\begin{equation}
I(\nu)_{\rm dust} = \epsilon B_\nu(T_{\rm dust}) 
	\left( \frac{\nu}{\nu_0} \right)^{\beta_d} 
\label{dust_eq}
\end{equation}
with $\beta_d \sim 1.7$.
CMB emission follows a Planck spectrum,
with power-law spectral index $\beta=2$ in the 
low-frequency Rayleigh-Jeans portion
and an exponential cutoff 
on the Wien side of the blackbody spectrum
(Fig \ref{foreground_fig}).

CMB emission can be distinguished from Galactic foregrounds
based on their different frequency spectra.
The number of independent frequency channels
must equal or exceed the number of free parameters
to be derived from a multi-frequency fit.
A conservative approach requires at least ten frequency channels
at the millimeter wavelengths 
where the CMB is brightest:
three parameters 
(amplitude, spectral index, and spectral curvature)
for synchrotron emission, 
three parameters 
(amplitude, spectral index, and frequency-dependent
polarization fraction) 
for thermal dust emission,
two parameters 
(amplitude and spectral index) 
for electric dipole emission from a population of rapidly spinning dust grains, 
one parameter 
(amplitude) 
for free-free emission (thermal bremsstrahlung) 
reflected from diffuse interstellar dust, 
and one parameter (amplitude) 
for CMB polarization.

%--------------------------------------------------------------------------
% Figure 7: Dust simulation
%--------------------------------------------------------------------------
\begin{figure}[t]
\centerline{
\includegraphics[height=3.0in]{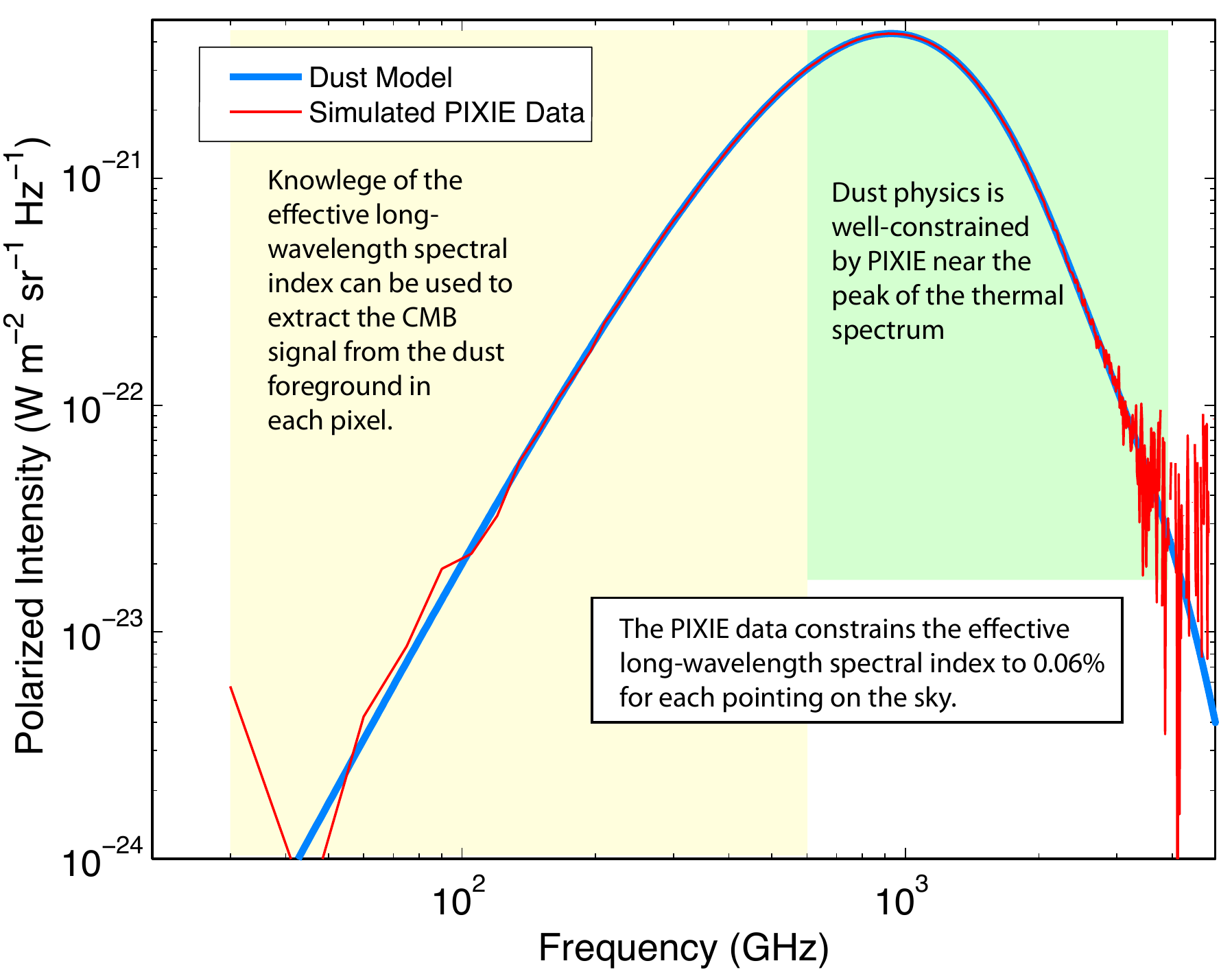}
}
\caption{
Foreground subtraction in the frequency range 60--600 GHz
where CMB emission is appreciable
is informed by dust physics constrained by 
PIXIE's high-frequency data.
Simulated PIXIE data are shown with a
two-component polarized foreground model.
A Levant-Marquandt minimization
recovers the long-wavelength dust spectral index
to accuracy $\pm 0.001$
and the CMB component to accuracy $\pm 1$ nK
within each independent pixel.
}
\label{dust_fig}
\end{figure}
%--------------------------------------------------------------------------

A commonly used technique forms a linear combination of frequency channels,
\begin{equation}
T_{\rm ILC} = \sum_\nu \alpha_\nu T_\nu ~,
\label{ilc_eq}
\end{equation}
to separate CMB from foreground emission based on the different spectra. 
If the component spectra are known, 
the coefficients $\alpha_\nu$
may be chosen such that a CMB spectrum is recovered with unit response 
while the foreground signals are canceled
\citep{bennett/etal:1992,
bennett/etal:2003}.
A useful way to express sensitivity is to quote the 
Foreground Degradation Factor (FDF), 
the ratio of the noise per pixel in the foreground-reduced map 
to corresponding noise obtained using straight noise weighting. 
PIXIE has many more frequency channels than foreground components, 
allowing efficient foreground suppression without excessive noise penalty. 
PIXIE achieves FDF=1.02 -- 
the noise penalty for rejecting foregrounds with PIXIE is only 2\%. 
This noise penalty is included in all estimates of CMB sensitivity.

The internal linear combination technique 
requires knowledge of the foreground spectral indices.
PIXIE has 400 effective channels from 30 GHz to 6 THz,
allowing independent determination of the spectral indices
within each  sky pixel.
PIXIE primarily observes at frequencies above 100 GHz 
where the dominant foreground is dust. 
Simulations using realistic polarized foregrounds 
\citep{finkbeiner/etal:1999,
hildebrand/kirby:2004}
and instrument noise
successfully recover the long-wavelength dust index 
from the input maps
to 0.06\% precision,
corresponding to $\pm 0.001$ uncertainty
in the dust spectral index
(Fig \ref{dust_fig}).
The resulting error in the recovered CMB component 
is below 1 nK.

\subsection{Systematic Errors}
Reliable detection of the primordial gravity-wave signal 
requires rigorous control of systematic errors.
PIXIE uses multiple levels of signal modulation
spanning 11 orders of magnitude in time
to reject systematic errors.

\begin{figure}[t]
\centerline{
\includegraphics[width=4.7in]{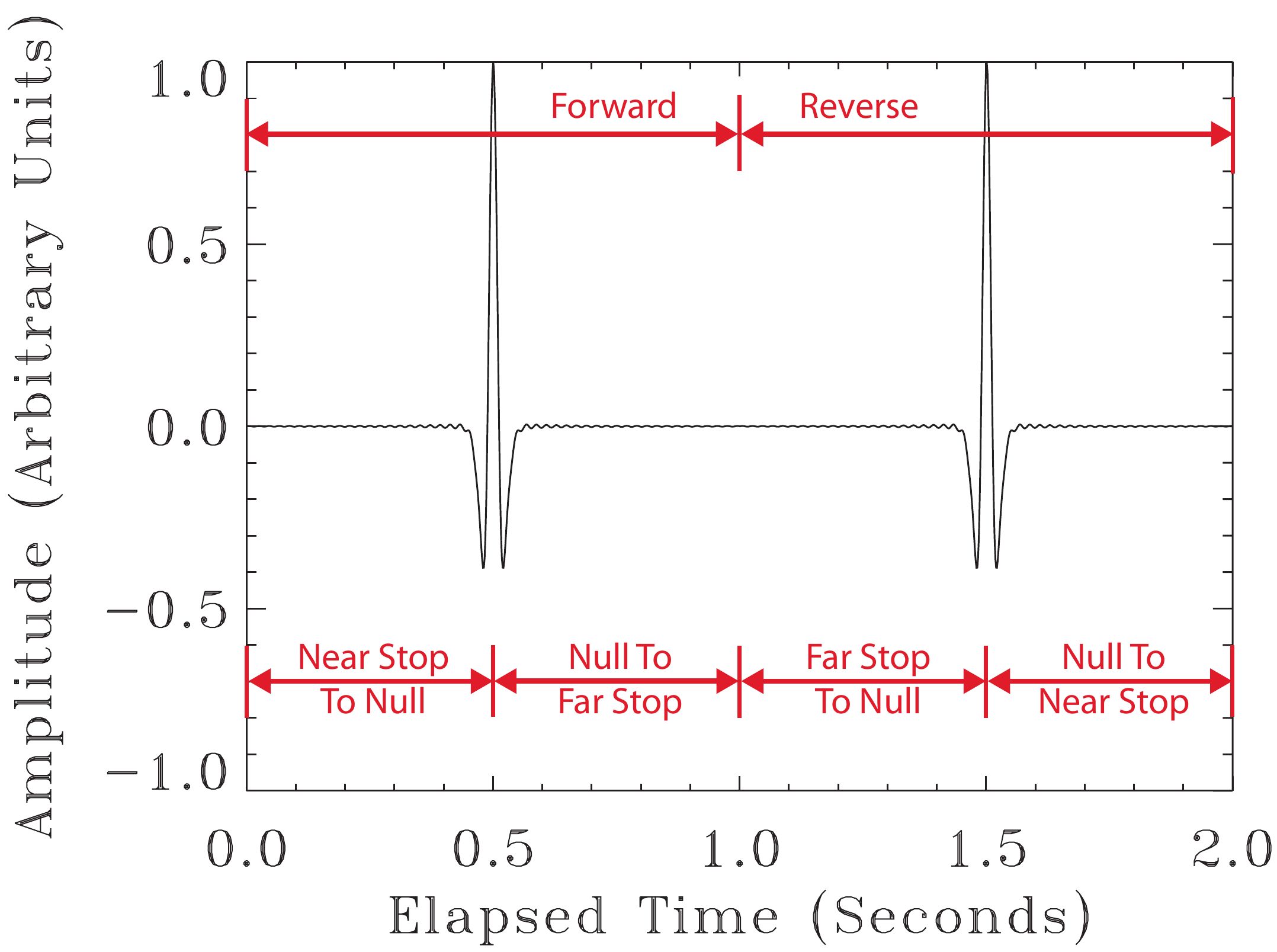}}
\caption{
Simulated fringe pattern from a single detector 
observing an unpolarized CMB source
with mirror period 2 sec.
Sky signals must follow multiple time-and space-reversal symmetries, 
allowing straightforward identification of instrumental signals.
}
\label{mtm_fringe}
\end{figure}

\begin{figure}[t]
\centerline{
\includegraphics[width=5.0in]{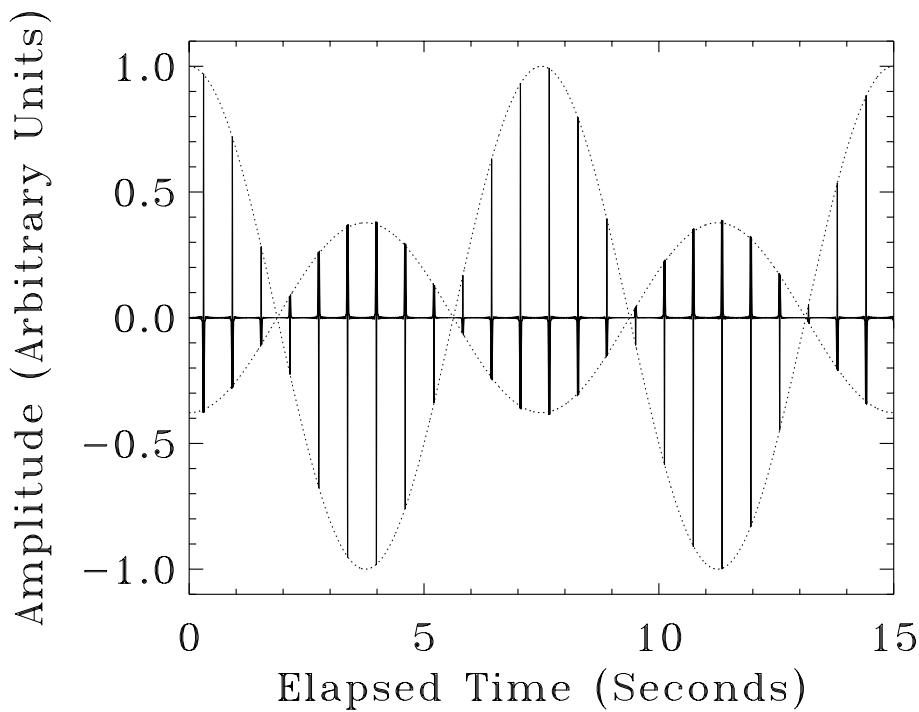}}
\caption{
Simulated fringe pattern from a polarized source 
for one complete spin period with mirror period 0.6 sec
and spin period 15 sec. 
The spacecraft spin imposes an amplitude modulation (dotted envelope) 
on the entire fringe pattern.
}
\label{amp_mod}
\end{figure}

\begin{itemize}

\item
{\bf Detector Readout (1 ms)}
An AC bias circuit modulates the detector output at 1 kHz
to reduce the effects of $1/f$ noise
in the detector or readout electronics.

\item
{\bf FTS Fringe Pattern (1 sec)}
The fringe pattern observed as the mirror moves
from one endpoint through the white-light null
to the opposite endpoint
modulates sky signals
on time scales from 1 ms to 1 s
(Fig \ref{mtm_fringe}).
True sky signals are independent of the sign
(near vs far side of null)
and direction
(forward vs back)
of the mirror motion,
allowing separation of sky signal
from instrumental effects.

\item
{\bf Spacecraft Spin (15 sec)}
Unlike simple polarization-sensitive detectors, 
where instrument rotation produces a sinusoidal response to polarized sky signals, 
PIXIE's rotation produces amplitude modulation
of the entire fringe pattern
at twice the spin frequency
(Fig \ref{amp_mod}).
The resulting modulated fringe pattern is readily distinguished
from a simple spin-locked sine wave or its harmonics,
suppressing spin-synchronous drifts. 
Sky signals are measured uniformly in azimuthal angle,
mitigating the effect of beam ellipticity and cross polar response.

\item
{\bf Calibration (3 hours)}
The calibrator changes position every even-numbered orbit
($\approx$3 hours),
reversing the sign of the fringe pattern from any sky signal.
The calibrator changes temperature every odd-numbered orbit. 
Observations with the calibrator warmer than the sky 
reverse the sign of the fringe pattern 
compared to observations with the calibrator colder than the sky.

\item
{\bf Orbit (2 days)}
The orbital scan re-visits each sky pixel 
for at least 20 consecutive orbits.
Pixels near the celestial poles are viewed every orbit.

\item
{\bf Orbit Precession (6 months)}
The observatory completes a full sky map every six months.
A sky pixel observed on the ascending node of the orbit 
is re-observed 6 months later on the descending node, 
allowing subtraction of orbital effects.

\end{itemize}

%--------------------------------------------------------------------------
% Figure 8: Fringe pattern
%--------------------------------------------------------------------------
%--------------------------------------------------------------------------

%--------------------------------------------------------------------------
% Figure 9: Amplitude modulation from spin
%--------------------------------------------------------------------------
%--------------------------------------------------------------------------

\noindent
In addition to the multiple levels of signal modulation,
PIXIE's highly symmetric design
provides additional rejection of potential systematic errors.

\begin{itemize}

\item
{\bf $x$--$y$ Symmetry}
The two detectors within each concentrator
share the same optical path
but observe orthogonal linear polarization states
(Eq. \ref{full_p_eq}).
Beam and pointing errors cancel to first order
when differencing the detectors.

\item
{\bf L--R Symmetry}
The $\hat{x}$ detector on the left side
measures the same sky signal as 
the $\hat{y}$ detector on the right side,
but with opposite sign
(Eq. \ref{full_p_eq}).
Beam and pointing errors cancel to first order
when summing these two detectors.

\item
{\bf A--B Symmetry}
The $\hat{x}$ detector on the left side
observes the same linear polarization
as the $\hat{x}$ detector on the right side,
but with the roles of the $A$ and $B$ beams reversed.
Differences reveal differential loss in the optics.

\item
{\bf Real--Imaginary Symmetry}
The spatial symmetry of the fringe pattern about zero path length
forces the sky signal entirely into the real part of the Fourier transform. 
The imaginary component of the Fourier transform 
provides an independent realization of the instrument noise,
including systematics and any $1/f$ component,
sampled at the same time and through the same optics 
as the noise in the sky spectra. 
The imaginary spectral maps may be analyzed identically 
to the real spectral maps (including multipole power spectra) 
as blind tests of systematics, 
with identical noise amplitude as the analyzed signal maps.

\end{itemize}

\begin{figure}[t]
\centerline{
\includegraphics[width=5.0in]{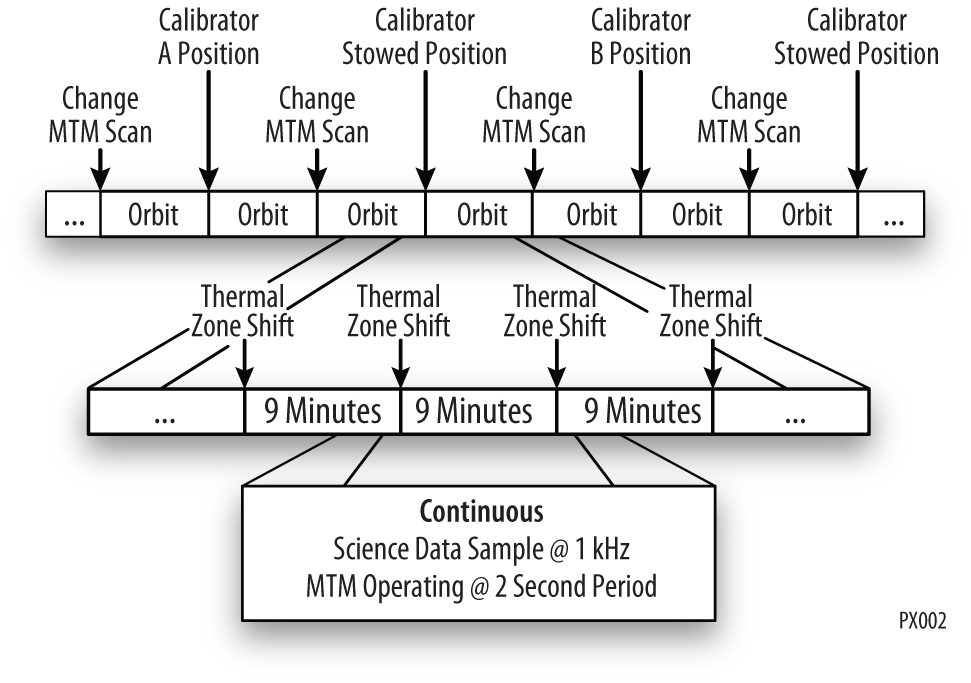}}
\caption{
Normal operations consist of a simple pre-set sequence
from an autonomous lookup table, which can be repeated indefinitely.
}
\label{ops_fig}
\end{figure}

\noindent
PIXIE operates in a continuous scanning mode,
sampling the detectors at $\sim$1 kHz
as the FTS mirror scan and spacecraft spin
modulate the sky signal.
Figure \ref{ops_fig} shows the nominal operation sequence.
Every 9 minutes the temperature of one of 20 optical surfaces
within the instrument is adjusted to a new value 
to measure its effective emissivity. 
Every other orbit the calibrator moves to a different position
(blocking beam A, blocking beam B, or stowed),
while the peak-to-peak mirror stroke length is adjusted 
to apodize the interferograms
(Appendix C).
These changes use an autonomous lookup table and 
require no direct ground intervention.
The observing strategy makes use of PIXIE's inherent symmetry
to further reduce potential systematic errors.

%--------------------------------------------------------------------------
% Figure 10: Operations
%--------------------------------------------------------------------------
%--------------------------------------------------------------------------

\begin{itemize}

\item
{\bf Null Operation}
PIXIE operates as a nulling polarimeter
to avoid the need for high-fidelity control 
of the relative bolometer calibration.
The $\hat{x}$ polarization from the A beam 
interferes with the $\hat{y}$ polarization from the B beam 
(Eq. \ref{full_p_eq}).
Only the \emph{difference} signal is modulated, 
reducing the requirement on gain match or gain drift 
by a factor of $10^5$.

\item
{\bf Isothermal Operation}
The entire telescope,
including the mirrors, FTS, support structure and enclosure,
is maintained within 10 mK of the sky temperature.
Rays that terminate within the instrument
(stray light)
simply replace a sky photon
with an equivalent photon
at nearly identical temperature,
reducing the requirement on stray light
or measured emissivities
by a factor of $10^4$.

\item
{\bf Thermal Stability}
PIXIE observes from a thermally stable sun-synchronous orbit 
to minimize thermal perturbations that might source systematic errors
(e.g. gain drifts). 
To further reduce thermal effects, 
the front-end analog electronics are housed 
in a thermally-controlled enclosure. 
Active thermal control and a long time constant 
reduce spin-locked temperature variation below 2 mK.

\item
{\bf Frequency Coverage}
PIXIE's frequency coverage extends well beyond the 
Wien cutoff in the CMB spectrum. 
Greybody emission from bright sources (Earth, Moon, Sun) 
in the far sidelobes
can be distinguished from CMB emission
based on color temperature alone.

\item
{\bf Sky Observations}
The detector sampling (1 ms), 
mirror stroke (1 sec), 
and spacecraft spin (15 sec) 
are fast compared to the 41 sec required 
for the beam to pass through a fixed spot on the sky.
PIXIE completes a measurement of the frequency spectrum 
of the Stokes parameters $I$, $Q$, and $U$
independently for each spin period at each position on the sky. 
Pixel-to-pixel comparisons are not required.

\end{itemize}

\noindent
A number of authors have examined systematic errors for CMB polarimetry
\citep{hu/etal:2003,		% Classification
carretti/etal:2004,		% Cross-polar response
bock/etal:2006,			% Weiss report
shimon/etal:2008,		% Beam asymmetry
mactavish/etal:2008,		% SPIDER
brown/etal:2009,		% Pointing, modulation
miller/etal:2009,		% 
takahashi/etal:2010,		% BICEP
bock/etal:2009,			% EPIC
odea/etal:2011}.		% SPIDER optics
These systematic errors may be grouped into several broad categories.
Instrumental drift or $1/f$ noise is mitigated by the FTS.
The Fourier transform acts on short stretches of data,
ranging from 330 ms to 1 second
depending on the mirror apodization stroke
(Table \ref{scan_table}).
Each interferogram is independent:
$1/f$ noise or baseline drifts 
on time scales much longer than 1 second
appear as a constant slope or low-order polynomial
in the spatial frequency basis of any single interferogram.
The Fourier transform of such low-order polynomials
affects only the lowest few spectral bins 
and does not project efficiently onto either the CMB or foreground spectra.
$1/f$ noise is thus inefficient at creating striping in 
the CMB polarization maps.

Scan-synchronous effects,
which accumulate coherently through multiple observations,
are of particular concern for CMB polarimetry.
Amplitude modulation of the observed fringe pattern by the spacecraft spin
efficiently rejects simple scan-synchronous pickup or offset variation.
Scan-synchronous modulation of the instrument responsivity 
(gain drift) 
is more serious.
Variation in responsivity at twice the spin period
will modulate the dominant unpolarized signal
to mimic a polarized source.
PIXIE's null design mitigates such a systematic error.
With the calibrator stowed,
the instrument responds only to polarized sky emission,
removing the source term for gain modulation.
Gain drifts only affect data when the calibrator is deployed.
The calibrator is maintained within 10 mK of the sky temperature
to allow operation near null.
The residual systematic error signal
is linear in the sky--calibrator temperature difference,
while true sky signals are independent of the calibrator.
Calibration data taken at different calibrator temperature
allows identification and removal of scan-synchronous gain drifts.
Residual modulation
(e.g. from instrument asymmetries)
can be rejected by comparing the signal from the 4 independent detectors.
Gain drifts are predominantly common mode
and produce identical signals on all 4 detectors,.
True sky signals have opposite sign
for specific detector pairs,
allowing separation from common-mode signals.

% -------------- Table 3: Systematic Errors --------------
\begin{table}[t]
{
\small
\label{syserr_table}
\begin{center}
\caption{Systematic Error Budget}
\begin{tabular}{| l | c | c | c | c | c | c | c | c | }
\hline 
Effect	& Leakage & \multicolumn{6}{ | c | }{PIXIE Mitigation} & Residual \\
\cline{3-8}
	&	  & FTS & Spin & Orbit & Xcal & Symmetry & Preflight & (nK) \\
\hline
Beam Ellipticity 	& $E \rightarrow B$ 
	& & \checkmark & \checkmark & & \checkmark & \checkmark & 2.7 \\
\hline
Cross Polarization 	& $\bigtriangledown T^2 \rightarrow B$ 
	& & \checkmark & & & \checkmark & \checkmark & 1.5 \\
\hline

Polarized Sidelobes	& $\Delta T \rightarrow B$
	& & \checkmark & \checkmark & &  \checkmark & \checkmark  & 1.1 \\
\hline
Instrumental Polarization & $\Delta T \rightarrow B$
	& & \checkmark & \checkmark & \checkmark & \checkmark & \checkmark  & $< 0.1$ \\
\hline
Polarization Angle	& $E \rightarrow B$
	& & & \checkmark & &  \checkmark & \checkmark & 0.7 \\
\hline
Pointing Offset		&  $\Delta T \rightarrow B$
	& & \checkmark & \checkmark & \checkmark & \checkmark & \checkmark & 0.7 \\
\hline
Relative Gain		& $\Delta T \rightarrow B$
	&\checkmark & \checkmark &   & \checkmark & \checkmark & & $< 0.1$ \\
\hline
Scan-Synch Gain		& $T \rightarrow B$
	&\checkmark & \checkmark &  & \checkmark & \checkmark & & $< 0.1$ \\	
\hline
Scan-Synch Offset	& $T \rightarrow B$
	& \checkmark & \checkmark &  & \checkmark & \checkmark & \checkmark & $< 0.1$ \\	 
\hline
\end{tabular}
\end{center}
}
\end{table}
%------------------------------------------------------------

Beam effects 
modulate the sky signal
and can produce systematic errors.
For example,
instrument rotation of an elliptical beam pattern
over an unpolarized sky
creates a spurious ``polarization'' response
at twice the spin frequency
as the elliptical beam
beats against the quadrupolar component of the local anisotropy.
Since beam effects directly modulate the true sky signal,
they can not be removed simply by altering the scan strategy
and are sometimes referred to as ``irreducible'' systematic errors.
The effect depends on anisotropy on angular scales
smaller than the instrument beam
and is thus of particular concern 
for an instrument like PIXIE
whose 2\ddeg6~beam diameter 
is larger than the degree angular scale 
of unpolarized CMB anisotropy.

%--------------------------------------------------------------------------
% Figure 11: Cross-polar systematic error map
%--------------------------------------------------------------------------
\begin{figure}[t]
\centerline{
\includegraphics[height=4.0in]{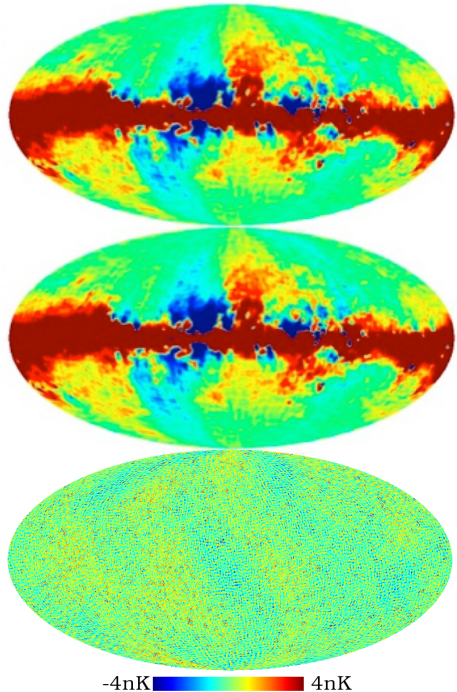}}
\caption{
Sky maps (Stokes $Q$ evaluated at 250 GHz)
from end-to-end systematic error simulations.
(top) Input sky model consisting of CMB and a 2-component dust model.
(middle) Output map from end-to-end simulations
including the modeled cross-polar beam response.
(bottom) Residual map
showing the systematic error from the cross-polar beam response.
}
\label{crosspol_fig}
\end{figure}
%--------------------------------------------------------------------------
PIXIE's differential design mitigates beam-related systematic errors.
With the calibrator stowed,
the unpolarized signal cancels to leading order,
leaving only the differential ellipticity
between the A- and B-side beams 
as a source of potential error.
Regardless of calibrator position,
polarized and unpolarized signals
produce a different sign on each of the four detectors
(Eq. \ref{diff_spectra_eq}),
allowing unambiguous separation
and rejection of beam effects.

A non-zero cross-polar beam response creates systematic errors
by mixing the Stokes $Q$ and $U$ parameters,
thereby aliasing 
the dominant E-mode polarization
into a spurious B-mode pattern.
The dominant cause is the quadrupolar component
of the cross-polar beam pattern\footnote{
A cross-polar response
that is uniform across the beam
affects the amplitude of the response to a
polarized sky signal 
(polarization efficiency)
but does not mix the $E$ and $B$ modes.
}.
PIXIE's non-imaging optics produce
a tophat beam
with nearly uniform polarization response,
mitigating this effect.
As with other beam-related effects,
true polarized sky signals
enter each detector
at a specific relative phase in spin angle
and can be identified by the appropriate detector-pair difference.

Table \ref{syserr_table} lists the major sources
of systematic error
and their estimated effect on the PIXIE polarization results.
We estimate the amplitude of each effect 
using end-to-end simulations.
Figure \ref{crosspol_fig} shows an example
for the cross-polar instrument response.
We generate a multi-frequency sky model
consisting of the CMB monopole,
temperature anisotropy,
and polarization
plus a relatively simple Galactic foreground consisting
of a 2-component dust model
\citep{finkbeiner/etal:1999}.
We assume a constant 2\% dust polarization fraction
and adopt dust polarization direction
from the WMAP K-band (synchrotron) data.
The top panel of Figure \ref{crosspol_fig}
shows the resulting sky model 
in the 250 GHz frequency channel.

We next `fly'' the mission over the simulated sky model,
using nominal instrument pointing and spin
to generate a set of time-ordered data
that includes the modeled cross-polar beam response.
An independent mapping pipeline
reads the time-ordered archive
and solves for the Stokes $I$, $Q$, and $U$ maps
within each synthesized frequency bin.
We then subtract the output sky maps
from the input model used to generate the simulation
to create a set of maps
showing the systematic error resulting from 
the cross-polar beam response.
The bottom panel of Figure \ref{crosspol_fig}
shows the cross-polar systematic error map 
in the 250 GHz frequency channel.
The cross-polar response creates a systematic error
with peak-to-peak amplitude of $\sim$4 nK,
with rms amplitude 1.5 nK averaged over the full sky.
Similar simulations provide estimates
for each systematic error term in Table \ref{syserr_table}.
Systematic errors are small compared to the instrument noise.

\section{Data Set and Science Goals}
PIXIE will map the full sky in absolute intensity and linear polarization
(Stokes $I$, $Q$, and $U$)
with angular resolution 2\ddeg6~
in each of 400 frequency channels
15 GHz wide
from 30 GHz to 6 THz.
The calibrated spectral data at each frequency
will be binned into 49152 sky pixels
each 0\ddeg9~ in diameter
using the HEALPIX pixelization
\citep{gorski/etal:2005}.
Typical sensitivities within each mid-latitude pixel are
\begin{equation}
\delta I_\nu^{I} = 4 \times 10^{-24}~
{\rm W~m}^{-2}~{\rm s}^{-1}~{\rm sr}^{-1}
\nonumber
\end{equation}
for Stokes $I$
and
\begin{equation}
\delta I_\nu^{QU} = 6 \times 10^{-25}~
{\rm W~m}^{-2}~{\rm s}^{-1}~{\rm sr}^{-1}
\nonumber
\end{equation}
for Stokes $Q$ or $U$.
The resulting data set supports a broad range of science goals.

\subsection{Inflation}
The primary science goal is the characterization
of primordial gravity waves from an inflationary epoch
through measurement of the CMB B-mode power spectrum.
PIXIE will measure the CMB linear polarization
to sensitivity of 70 nK per $1\deg \times 1\deg$ pixel,
including the penalty for foreground subtraction.
Averaged over the cleanest 75\% of the sky,
PIXIE can detect B-mode polarization
to 3 nK sensitivity,
well below the 30 nK predicted from large-field inflation models.
The sensitivity is comparable to the ``noise floor'' 
imposed by gravitational lensing,
and allows robust detection of primordial gravity waves
to limit $r < 10^{-3}$ at more than 5 standard deviations.

\begin{figure}[t]
\centerline{
\includegraphics[width=5.0in]{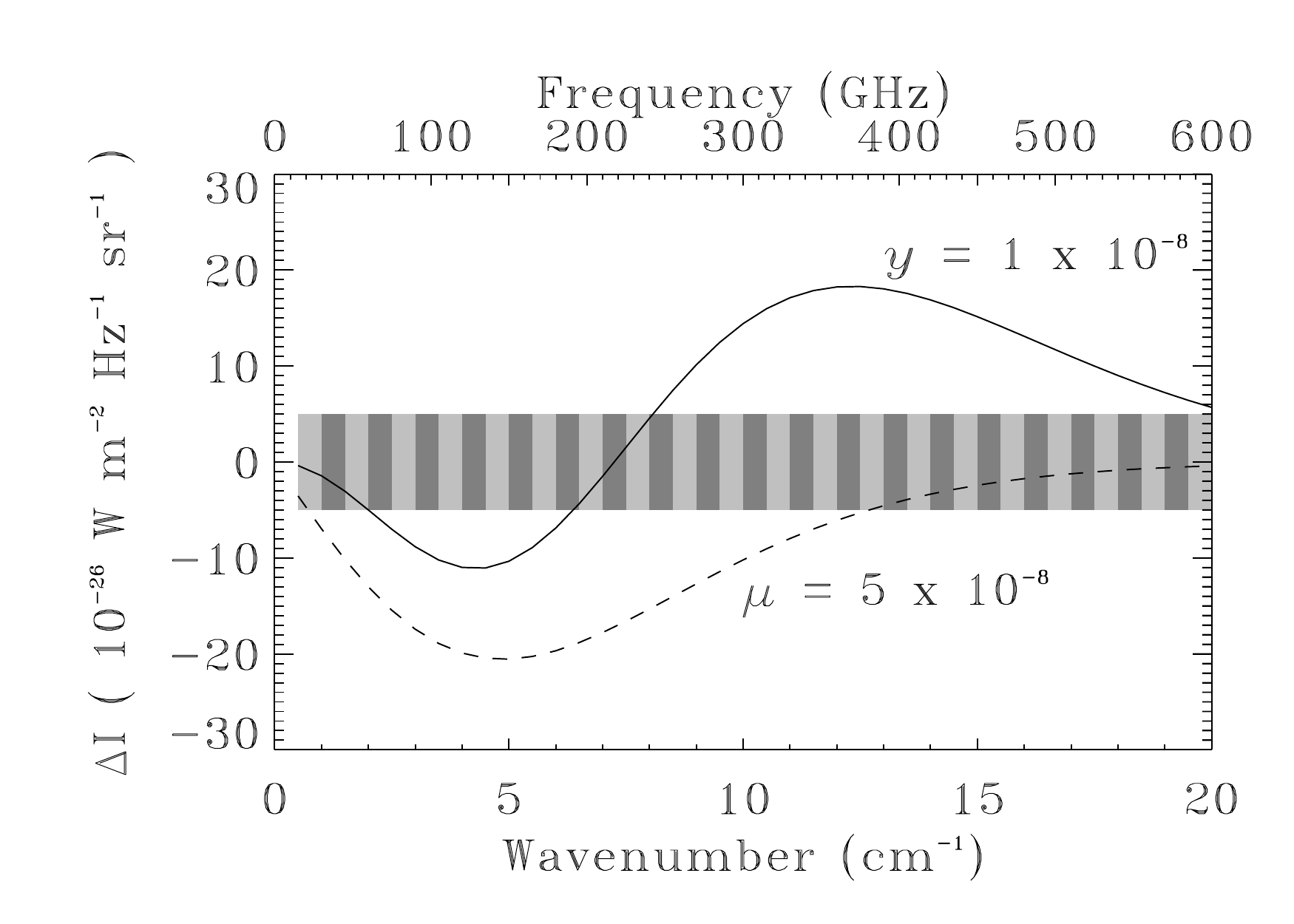}}
\caption{
Distortions to the CMB blackbody spectrum
compared to the PIXIE instrument noise in each synthesized frequency channel.
The curves show 5$\sigma$ detections
of Compton ($y$) 
and chemical potential ($\mu$)
distortions.
PIXIE measurements of the $y$ distortion
determine the temperature of the intergalactic medium
at reionization,
while the $\mu$ distortion
probes early energy release
from dark matter annihilation
or Silk damping of primordial density perturbations.
}
\label{spectrum_fig}
\end{figure}

PIXIE also constrains inflationary models 
through measurements of distortions in the CMB blackbody spectrum. 
Inflation generates density fluctuations on all physical scales. 
The largest scales freeze out 
and later re-enter the particle horizon 
as the fluctuations observed by WMAP and Planck. 
On smaller scales, photon diffusion (Silk damping) 
erases the primordial fluctuations
and transfers their energy to the radiation bath.
Energy injected to the CMB distorts its spectrum from a blackbody. 
For energy injected at redshift
$10^4 < z < 10^7$,
the distorted spectrum is characterized by a chemical potential
\begin{equation}
\mu = 1.4 \frac{\Delta E}{E}
\label{mu_eq}
\end{equation}
proportional to the fractional energy release $\Delta E/E$
to the CMB
\citep{burigana/etal:1991,
daly:1991,
hu/scott/silk:1994}.
PIXIE compares the (monopole) CMB spectrum 
to a blackbody calibrator with $\mu$K precision,
corresponding to a chemical potential
$\mu < 10^{-8}$
(Fig. \ref{spectrum_fig}).
Constraints on chemical potential distortions
in the CMB spectrum probe the amplitude of matter fluctuations
down to physical scales as small as 1 kpc (1 solar mass).

%--------------------------------------------------------------------------
% Figure 12: Spectral Distortions
%--------------------------------------------------------------------------
%--------------------------------------------------------------------------

\subsection{Dark Matter}
The chemical potential of the CMB spectrum
provides a limit to any early energy release.
Neutralinos are an attractive candidate for dark matter;
the annihilation of $\chi \bar{\chi}$ pairs
in the early universe 
leads to an observable distortion in the CMB.
The chemical potential 
can be estimated as
\begin{equation}
\mu \sim 3 \times 10^{-4} 
~f
~\left( \frac{\sigma v}{6 \times 10^{-26}~{\rm cm}^3~{\rm s}^{-1}} \right)
~\left( \frac{m_\chi}{1 ~{\rm MeV}} \right)^{-1}
~\left( {\Omega_{\chi}}h^2 \right)^2 .
\label{wimp_mu_eq}
\end{equation}
where
$f$ is the fraction of the total mass energy
released to charged particles,
$\langle \sigma v \rangle$ 
is the velocity-averaged annihilation cross section,
$\Omega_{\chi}$ is the dark matter density,
and
$h = H_0 / 100~{\rm km~s}^{-1}~{\rm Mpc}^{-1}$
is the Hubble constant
\citep{silk/stebbins:1983,
mcdonald/etal:2001}.
The dark matter annihilation rate varies as 
the square of the number density.
For a fixed $\Omega_{\chi}$
the number density is inversely proportional 
to the particle mass.
The chemical potential distortion
is thus primarily sensitive to lower-mass particles.
PIXIE will probe neutralino mass range $m_\chi \lsim 80$ keV
to provide a definitive test
for light dark matter models
\citep{devega/sanchez:2010}.

Neutralino models are only one class of potential dark matter candidates.
Many supersymmetric models predict that 
the lightest stable supersymmetric particle 
is the gravitino, the superpartner of the graviton, 
rather than the photino (or more generally the neutralino), 
the superpartner of the photon.  
Such superWIMP models can yield dark matter densities near the observed value
while suppressing dark matter interactions
to values well below the weak scale.
For dark matter experimentalists, 
these superWIMP models are a depressing possibility 
as the gravitino will be undetectable 
in any foreseeable terrestrial dark matter search.

%--------------------------------------------------------------------------
% Figure 13: Chemical potential from slepton decay
%--------------------------------------------------------------------------
\begin{figure}[t]
\centerline{
\includegraphics[width=5.0in]{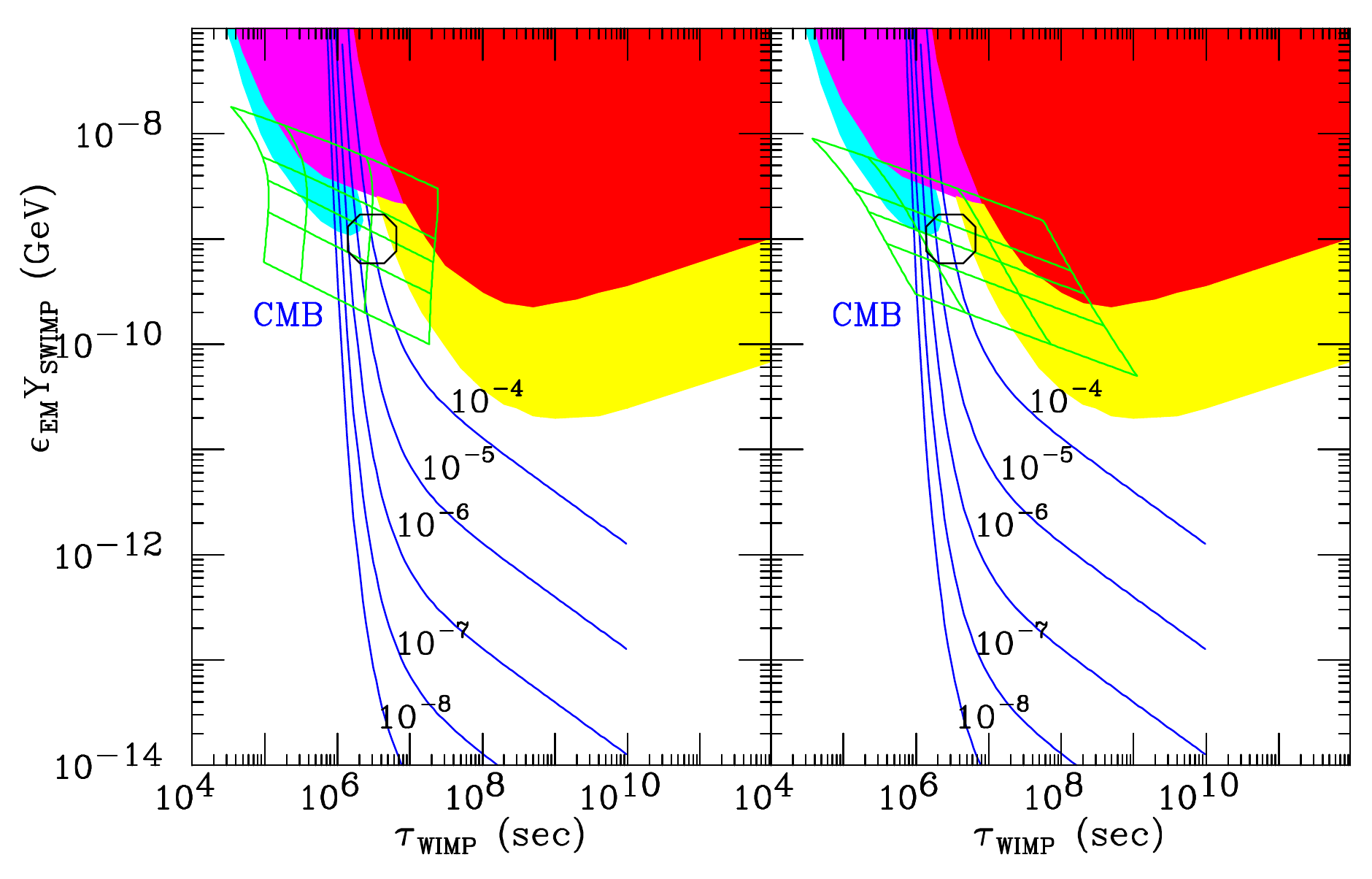}}
\caption{Chemical potential distortion to the CMB
from slepton WIMP $\rightarrow$ gravitino superWIMP decay,
compared to current bounds on
the lifetime $\tau$ 
and electromagnetic energy deposit $\epsilon_{\rm EM} Y$
for 
$\tilde{B}$ (left) and
$\tilde{\tau}$ (right) models
(adapted from \citep{feng/etal:2003}).
Shaded regions are excluded by Big Bang nucleosynthesis constraints.
The grid shows the preferred region of parameter space
where the relic density of gravitino superWIMPS 
matches the observed dark matter density.
Grid values show mass 
$m_{\rm SWIMP}$ = 100, 300, 500, 1000, and 3000 GeV
(top to bottom) 
and mass difference
$m_{\rm WIMP} - m_{\rm SWIMP}$ = 600, 400, 200, and 100 GeV
(left to right).
The octagon shows the region
required to reduce $^7$Li to observed values.
The PIXIE limit $\mu < 10^{-8}$
probes a critical region of parameter space.
}
\label{slepton_fig}
\end{figure}
%--------------------------------------------------------------------------

CMB spectral distortions offer a novel probe of supersymmetric physics.
The decay of heavier supersymmetric leptons 
to gravitinos in the early universe
also releases energy to Standard Model particles,
distorting the CMB blackbody spectrum
\citep{feng/etal:2003}.
Figure \ref{slepton_fig}
shows the amplitude of the resulting chemical potential $\mu$
as a function of the lifetime $\tau$
and electromagnetic energy release $\epsilon_{\rm EM} Y$
for slepton WIMP $\rightarrow$ gravitino superWIMP decay.
PIXIE explores an important region of superWIMP parameter space
not excluded by any other bound,
offering the potential to detect one of the most elusive
dark matter candidates.

\subsection{Reionization}
The ionization history of the universe is our most direct probe 
of the star formation history at redshifts greater than 7. 
Theoretical studies
\citep{barkana/loeb:2007,
meiksin:2007}
suggest our universe may have had a complex star formation history: 
this signature of the extended ionization history 
is traced by the E-mode polarization power spectrum 
on angular scales
$\theta > 20\deg$.
In some scenarios
\citep{ricotti/etal:2008},
early black holes produce low levels of ionization 
at redshift $z > 50$,
while in other scenarios,
metal-free Pop III stars partially ionize the universe 
and suppress further star formation until
a delayed second generation of stars finally ionizes the universe
\citep{cen:2003}.
While Planck data will improve existing constraints on the optical depth,
Planck is noise-limited at multipoles
$15 < \ell < 30$,
the range most sensitive to the details of the star formation history,
and can thus test only the most extreme models
\citep{mukherjee/liddle:2008}.
PIXIE will make a cosmic variance limited measurement
of the E-mode power spectrum at $\ell < 30$
to characterize the ionization history.

Early stars not only ionize the universe but also heat the gas. 
Microwave background photons inverse Compton scatter off this hot gas 
to produce a distinctive distortion in the CMB blackbody spectrum. 
The amplitude of this Compton distortion 
is characterized by the parameter
\begin{equation}
y = \int \frac{k [T_e(z) - T_r(z)]}{m_e c^2} n_e(z) \sigma_T c ~dz,
\label{compton_eq}
\end{equation}
where
$m_e$, $n_e$, and $T_e$
are the electron mass, number density, and temperature,
$\sigma_T$ is the Thomson cross section,
$T_r$ is the CMB radiation temperature,
and $k$ is Boltzmann's constant
\citep{zeldovich/sunyaev:1969}.
The resulting distortion
depends on both the
optical depth $\tau$
and gas temperature $T_e$,
\begin{equation}
y = 2 \times 10^{-7} 
~\left( \frac{\tau}{0.1} \right) 
~\left( \frac{T_e}{10^4~{\rm K}} \right) .
\label{y_reion}
\end{equation}
PIXIE measures the Compton spectral distortion 
to limit $y < 2 \times 10^{-9}$.
In combination with PIXIE's measurement of $\tau$
(observed through the same optics
and using the same calibration),
PIXIE will determine the temperature of the intergalactic medium
to $\sim$5\% precision at redshift $z \sim 11$.

The mean temperature of the universe
depends sensitively on the ionizing spectrum of the first objects.
PIXIE's precision measurement of the electron temperature 
will distinguish whether the first stars were
``classical'' Pop II stars
or more exotic objects such as very massive Pop II stars
or black holes.
No other measurement has such power to determine the nature
of the first luminous objects in the universe.

%--------------------------------------------------------------------------
% Figure 14: Dust physics
%--------------------------------------------------------------------------
\begin{figure}[t]
\centerline{
\includegraphics[width=5.0in]{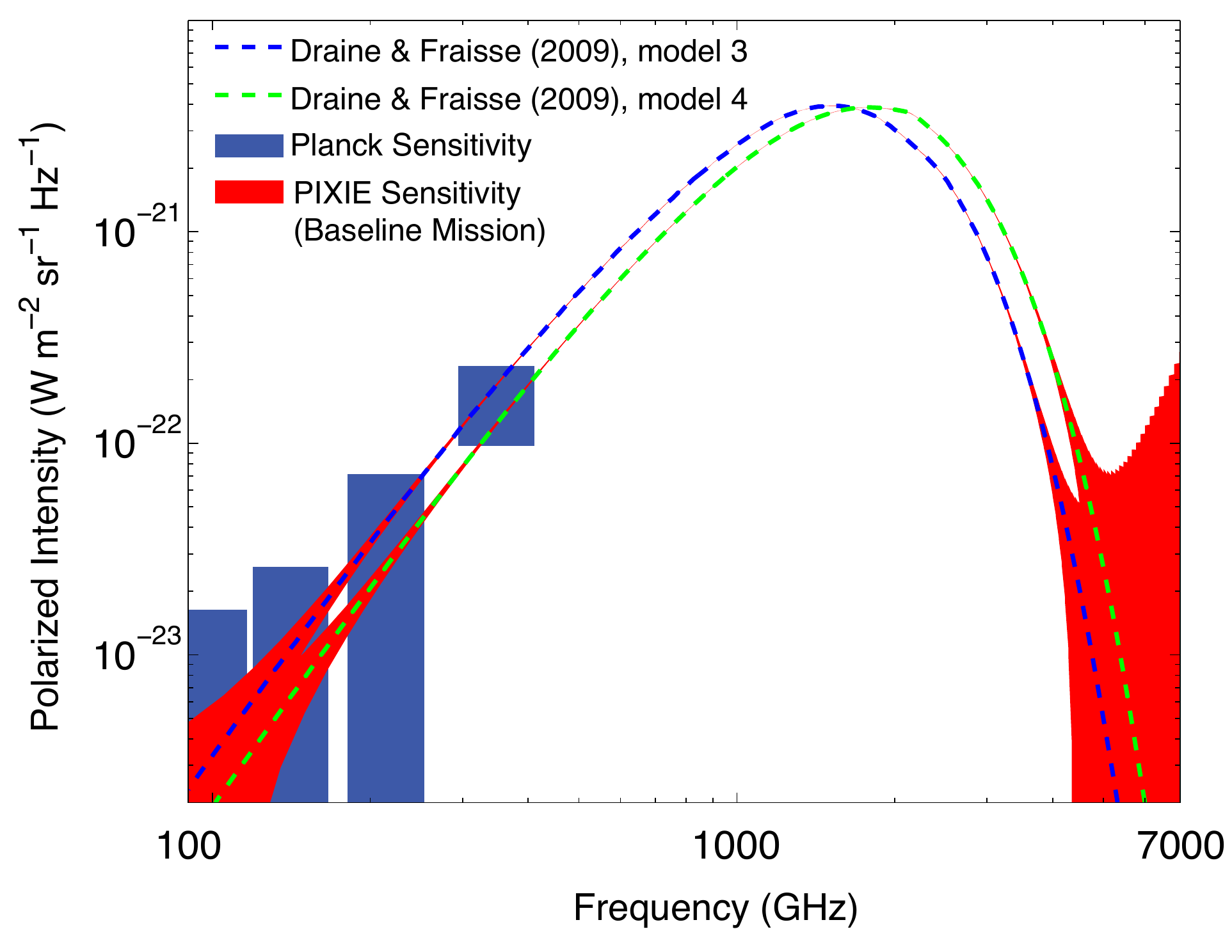}}
\caption{
PIXIE unambiguously separates the observed dust emission 
to determine the chemical composition of the diffuse dust cirrus 
across the entire sky. 
The figure shows the dust fractional polarization
for models with only silicate contribution (blue curve) 
or both silicate and carbonaceous dust (green curve)
\citep{draine/fraisse:2009}. 
For comparison, the Planck sensitivity 
for the four highest frequency polarized channels is also shown.
}
\label{dust_chem_fig}
\end{figure}
%--------------------------------------------------------------------------

\subsection{Star Formation History}
The cosmic infrared background (CIB) 
results from cumulative emission from dusty galaxies
since their epoch of formation.
It consists of X-ray, UV, and optical emission 
from stars and active galactic nuclei,
absorbed by dust and re-radiated at infrared wavelengths.
It therefore provides an integral constraint 
on the cosmic star formation history of the universe. 
The CIB arises primarily from galaxies with redshifts between 1 and 2
due to the steeply rising dust spectrum, 
which cancels the cosmological dimming due to redshift.  
The absolute emission spectrum as a function of angular scale 
is sensitive to the entire history of dusty star-forming galaxies
\citep{knox/etal:2001}.
Spatial fluctuations in the CIB should correlate
with the integrated Sachs-Wolfe effect
and gravitational lensing fluctuations
\citep{cooray/etal:2010}.
The combination is a powerful probe of the
characteristic mass of star-forming galaxies
and the integrated cosmic star formation history.
PIXIE's sensitivity at large angular scales
over a broad range of frequencies
spanning the peak CIB intensity
will complement measurements by Herschel, Planck,
and other observatories.

\subsection{Interstellar Medium}
PIXIE's all-sky spectral cubes
provide important insights into the interstellar medium (ISM)
within the Galaxy.
Thermal emission from dust within our galaxy 
dominates the far-infrared sky brightness. 
Dust emission is partially polarized by magnetic alignment of aspherical grains
\citep{lazarian:2007}.
While there are appealing theories 
detailing the alignment mechanism 
and the composition of the aligned grain populations,
there is minimal observational evidence
available to constrain these theories. 
PIXIE has both the sensitivity and frequency coverage 
to distinguish different models
of the polarized intensity based on their chemical composition.

%--------------------------------------------------------------------------
% Figure 15: Line emission
%--------------------------------------------------------------------------
\begin{figure}[t]
\centerline{
\includegraphics[width=5.0in]{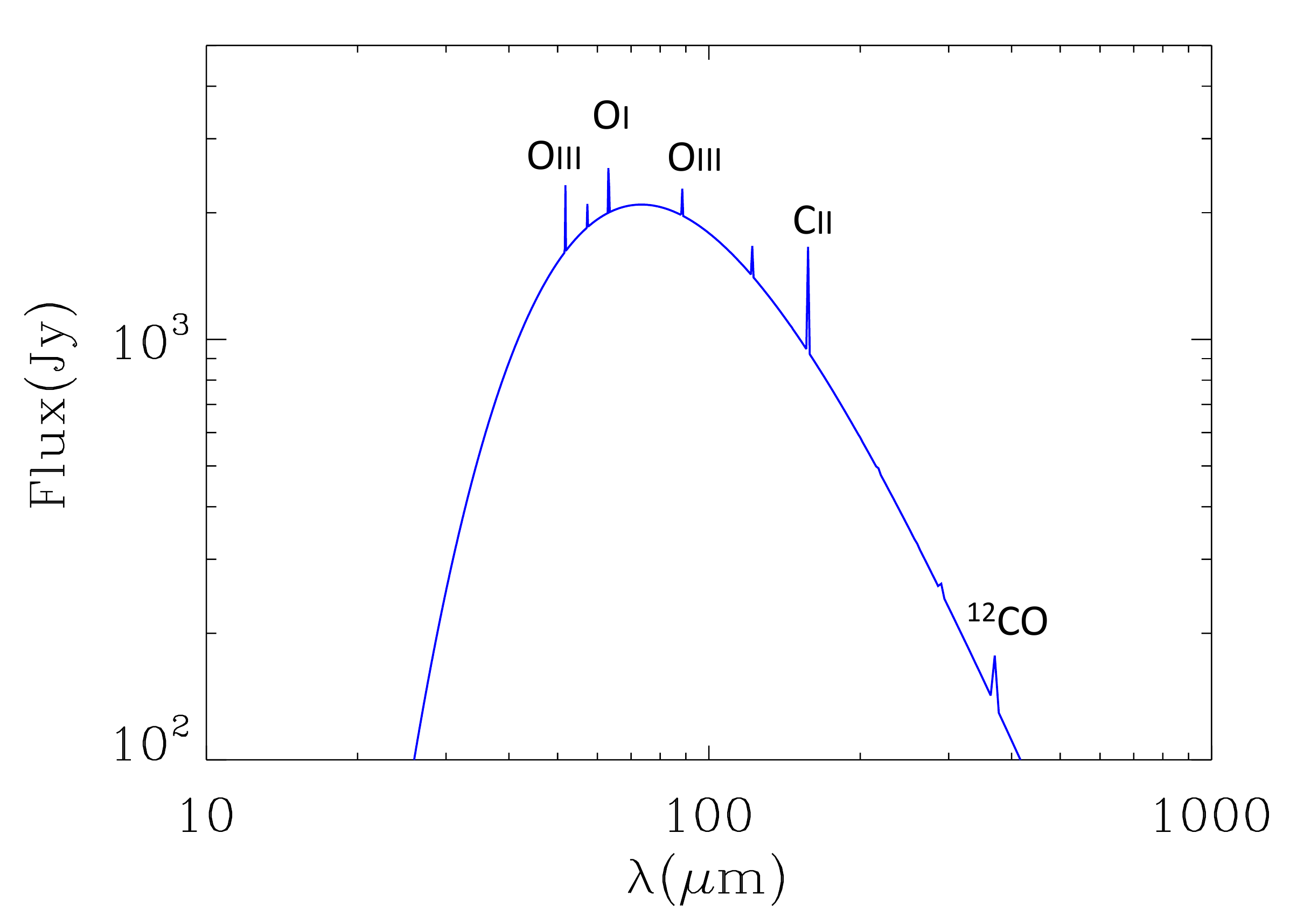}}
\caption{
Herschel observations of the thermal dust and line emission spectrum 
from the starburst galaxy M82,
convolved to the 15 GHz (300 km s${-1}$) 
resolution of the PIXIE instrument. 
PIXIE will provide similar spectra along lines of sight 
to large star forming complexes and molecular clouds along the galactic plane.
}
\label{m82_spectrum_pixie}
\end{figure}
%--------------------------------------------------------------------------

Figure \ref{dust_chem_fig}
illustrates the sensitivity to dust properties.
The polarized intensity of dust emission
depends on the chemical composition
(silicate vs carbonaceous).
PIXIE maps dust polarization at frequencies spanning the peak emission
to determine the chemical composition.
The resulting full-sky map of 
the chemical composition of the diffuse dust cirrus
provides important insights into the life cycle
of dust creation and destruction in the interstellar medium.
Physically motivated models of dust emission 
are also an important tool 
to subtract the dust foreground from maps of primordial CMB polarization
({\S}3.2).

An important driver of Galactic evolution 
is the chemical enrichment of the ISM 
in heavy elements, molecules, and dust. 
This enrichment is the result of the cumulative contribution 
of stellar nucleosynthetic processes 
over the star formation history of the Milky Way.  
The 50~$\mu$m to 1~cm spectral region covered by PIXIE 
contains important lines from atomic, ionized and molecular species, 
as well as continuum emission from the dust and gas in the ISM. 
PIXIE observations will thus provide important information 
for determining the abundances and composition of key elements 
and the dust in the ISM, the physical conditions of the host ISM phases, 
and the spatially resolved star formation rate in the Galaxy. 

The brightness of the far IR sky is dominated by 
thermal emission from dust. 
Figure \ref{m82_spectrum_pixie} 
depicts the Herschel spectrum of  the starburst galaxy M82, 
convolved to the 15 GHz resolution of the PIXIE instrument.  
PIXIE will provide similar spectra along the line of sight 
to large star forming complexes and molecular clouds along the galactic plane. 
Table \ref{line_table} 
lists selected emission lines 
from ionized, neutral and molecular species in the ISM.
The lines are organized according to the four major gas phases of the ISM 
from which they arise: 
\begin{itemize}
\item {\bf hot coronal gas}, characterized by densities and temperatures, 
\{n, T\} = \{$\sim 0.004$~cm$^{-3}$, $\gtrsim 10^{5.5}$~K\},  
collisionally ionized and maintained by expanding supernova blast waves; 
\item {\bf HII regions} with 
\{n, T\} = \{$\sim 1-10^4$~cm$^{-3}$, $\sim 10^4$~K\}, 
photoionized by massive OB stars; 
\item {\bf molecular clouds}, with 
\{n, T\} = \{$\sim 10^{2-6}$~cm$^{-3}$, $\sim 10-50$~K\}, 
heated by cosmic rays and photoelectrons from dust; and 
\item {\bf photodissociation regions (PDRs)} 
which are the interface between the molecular clouds and the ionized gas. 
Their characteristic density and temperatures are 
\{n, T\} = \{$\sim 1-10^4$~cm$^{-3}$, $\sim 10-5000$~K\}, 
and they are heated by photoelectrons from dust and cosmic rays. 
\end{itemize}

% -------------- Table 4: Line Emission --------------
\begin{table}[t]
{
\small
\caption{Selected Line Emission From the Interstellar Medium}
\label{line_table}
\begin{center}
\begin{tabular}{| l  l | l  l | l  l | l  l |}
\hline 
\multicolumn{2}{| c |}{Molecular} &
\multicolumn{2}{| c |}{Photodissociation} &
\multicolumn{2}{| c |}{H{\sc ii}} &
\multicolumn{2}{| c |}{Hot ($T > 10^5$ K)} \\
\multicolumn{2}{| c |}{Gas} &
\multicolumn{2}{| c |}{Regions} &
\multicolumn{2}{| c |}{Regions} &
\multicolumn{2}{| c |}{Gas} \\
\hline
CO $1 \rightarrow 0$ 	& 115 GHz &
Fe{\sc ii}		& 51.3, 87.4 $\mu$m &
Fe{\sc iii}		& 51.7 $\mu$m &
O{\sc iii}		& 51.8 $\mu$m \\
\hline
CO $2 \rightarrow 1$ 	& 231 GHz &
Fe{\sc i}		& 54.3, 111.2 $\mu$m &
N{\sc iii}		& 57.3 $\mu$m &
N{\sc iii}		& 57.3 $\mu$m \\
\hline
CO $30 \rightarrow 29$ 	& 3438 GHz &
S{\sc i}		& 56.3 $\mu$m &
Fe{\sc ii}		& 87.4 $\mu$m &
Fe{\sc v}		& 70.4 $\mu$m \\
\hline
H$_2$O 			& 22 GHz &
O{\sc i}		& 63.2 $\mu$m &
Fe{\sc iii}		& 105.4 $\mu$m &
O{\sc iii}		& 88.4 $\mu$m \\
\hline
H$_2$O 			& 183 GHz &
Si{\sc i}		& 68.5, 129.7 $\mu$m &
N{\sc ii}		& 121.9 $\mu$m &
\multicolumn{2}{|c|}{} \\
\hline
CS $1 \rightarrow 0$ 	& 49 GHz &
O{\sc i}		& 145.5 $\mu$m &
Si{\sc i}		& 129.7 $\mu$m &
\multicolumn{2}{|c|}{} \\
\hline
CS $2 \rightarrow 1$ 	& 98 GHz &
C{\sc ii}		& 157.7 $\mu$m &
N{\sc ii}		& 205.2 $\mu$m &
\multicolumn{2}{|c|}{} \\
\hline
CS $4 \rightarrow 3$ 	& 196 GHz &
C{\sc i}		& 370.4, 609.1 $\mu$m &
\multicolumn{2}{|c|}{}  &
\multicolumn{2}{|c|}{} \\
\hline
\end{tabular}
\end{center}
}
\end{table}
%------------------------------------------------------------

\begin{figure}[t]
\centerline{
  \includegraphics[width=3.0in]{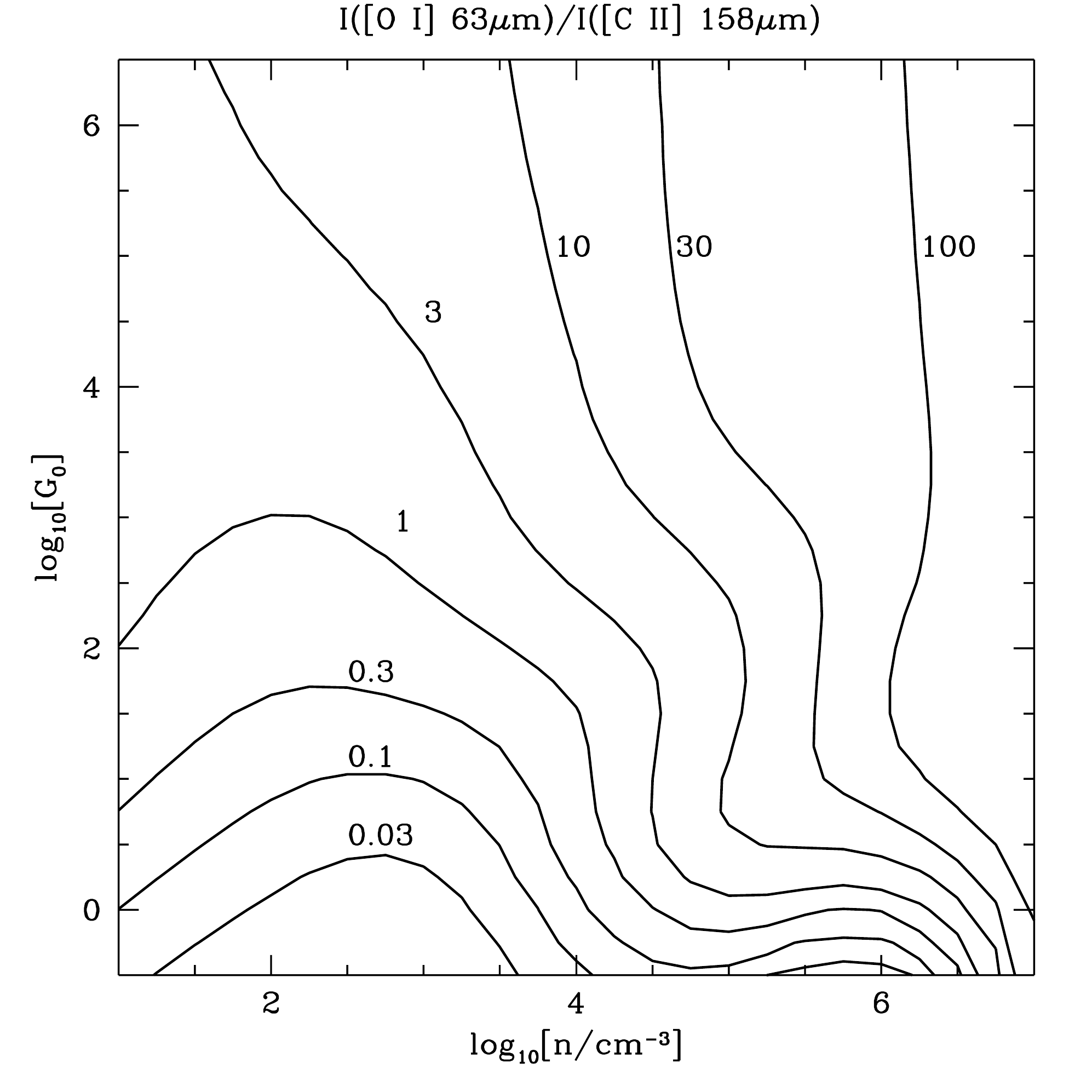}
  \includegraphics[width=3.0in]{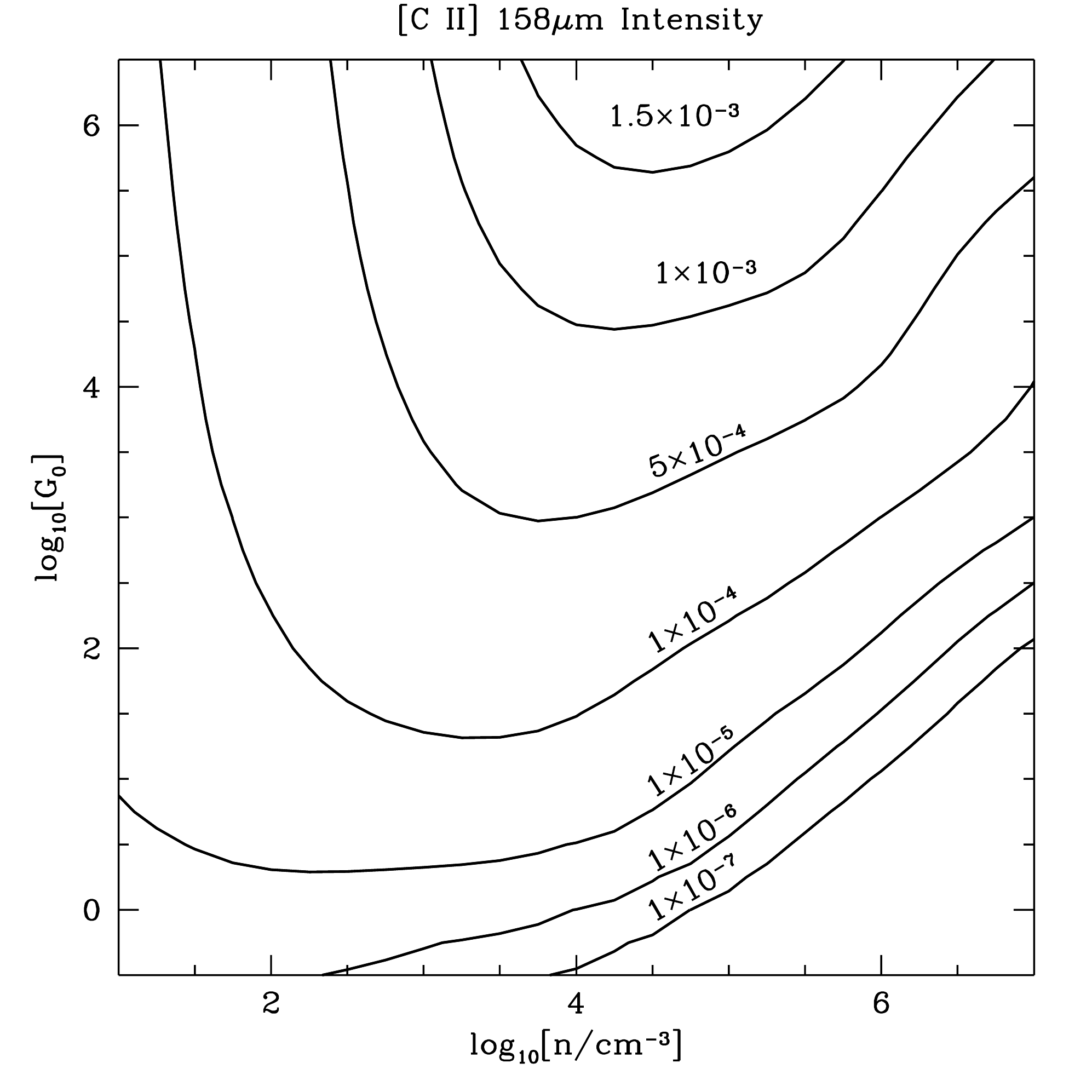}} 
\caption{
Contour diagrams, taken from \citet{kaufman/etal:1999},
illustrate the use of line ratios 
to determine the density $n$ and UV flux intensity $G_0$ in PDRs.
}
\label{line_ratio_fig}
\end{figure}

The different phases of the ISM cool by line emission 
and by continuum emission from dust and gas. 
In PDRs, the ionization state and intensity of the line emission 
are determined by the UV flux incident on the PDR, 
by the density of the gas, 
and since the gas is heated by photoelectrons 
ejected from the surface of dust grains, 
by the composition and size distribution of the dust as well. 
With its broad wavelength coverage,
PIXIE will provide all the necessary information to completely characterize
this important phase of the ISM. 
Line ratios from different excitation levels of the same species
serve as important density or temperature diagnostics of the gas.
Important density diagnostics include line ratios
[Fe{\sc {\sc i}{\sc i}}] 51~$\mu$m/87~$\mu$m,
[Fe{\sc i}] 54~$\mu$m/111 $\mu$m; 
[N{\sc {\sc i}{\sc i}}] 122~$\mu$m/205~$\mu$m; 
[O{\sc {\sc {\sc i}{\sc i}}{\sc i}}] 52~$\mu$m/88~$\mu$m; 
[O{\sc i}] 146~$\mu$m/63~$\mu$m; 
and [N{\sc {\sc i}{\sc i}}] 205~$\mu$m/122~$\mu$m.
Temperature diagnostics include line ratios
[N{\sc {\sc {\sc i}{\sc i}}{\sc i}}] 57~$\mu$m/[N{\sc {\sc i}{\sc i}}] 122~$\mu$m, 205~$\mu$m;
and [C{\sc {\sc i}{\sc i}}] 157~$\mu$m/[C{\sc i}] 370~$\mu$m, 609~$\mu$m.
Figure \ref{line_ratio_fig} 
depicts how the intensity of the UV radiation field incident 
on the surface of the PDR, 
$G_0$, 
can be used together with different line ratios 
to determine the elemental abundances 
and physical conditions of the PDRs that give rise to the line emission
\citep{kaufman/etal:1999}.

The density and ionization sources of Galactic H{\sc ii} regions 
can be completely characterized by 
their free-free emission and IR spectrum, 
both observed by PIXIE
through the same optics and with the same calibration.
The free-free emission determines the
density dependent rate of UV photons 
required to keep the region ionized. 
The spectrum of the dust emission is sensitive to 
metallicity and density of the H{\sc ii} region. 
The metallicity of the  H{\sc ii} regions can in turn 
be derived from their characteristic line emission. 
By creating template IR spectra for different 
densities, metalicities,and dust models, 
we may characterize the dust composition, 
gas density, 
and ionizing stellar sources that form the H{\sc ii} regions. 
Combined with the thermal dust emission arising from the neutral H{\sc i} gas
this information will be used to derive the star formation rate along
the different lines of sight in the Milky Way.

%--------------------------------------------------------------------------
% Figure 16: Line ratios
%--------------------------------------------------------------------------
%--------------------------------------------------------------------------

The characterization of physical conditions 
within the different gas phases of the ISM 
requires the decomposition of the continuum IR emission 
into its different gas phase components. 
Using the free-free emission maps as proxies for H{\sc ii} regions,
molecular lines as proxies for the molecular gas, 
and ancillary Galactic 21~cm surveys to trace the atomic gas, 
PIXIE observations will be correlated with
the emissions from these different ISM component. 
This will enable the decomposition of the thermal dust emission, 
providing the first detailed determination of the
temperature distribution, the abundance and composition of the dust 
in the different phases of the ISM. 

\section{Summary}

The Primordial Inflation Explorer 
will measure the intensity and linear polarization 
of the cosmic microwave background
to search for the predicted signature of gravity waves
excited during an inflationary epoch in the early universe.
It uses a polarizing Michelson interferometer
configured as a nulling polarimeter
to measure the inflationary signal
to the limits imposed by astrophysical and cosmological foregrounds.

PIXIE
achieves the sensitivity,
frequency coverage,
and control of systematic errors 
needed to characterize the inflationary signal
while remaining within the resources of NASA's Explorer program.
Multi-moded optics achieve background-limited sensitivity
using only 4 semiconductor bolometers.
A Fourier Transform Spectrometer
provides 400 synthesized frequency channels each 15 GHz wide
from 30 GHz to 6 THz,
yielding unprecedented ability to distinguish cosmic from Galactic emission
based on their difference frequency spectra.
The highly symmetric instrument design
and multiple layers of signal modulation
spanning 11 orders of magnitude in time
provide robust discrimination against systematic errors.
Detailed simulations of the entire mission,
including data analysis,
foreground subtraction,
and systematic error contribution,
demonstrate sensitivity $r < 10^{-3}$
at 5 standard deviations
over the cleanest 75\% of the sky.

PIXIE will measure the absolute intensity and linear polarization
(Stokes $I$, $Q$, and $U$ parameters)
over the full sky
at 2\ddeg6 ~angular resolution,
with typical sensitivity
$4 \times 10^{-24}~
{\rm W~m}^{-2}~{\rm s}^{-1}~{\rm sr}^{-1}$
for unpolarized emission
and 
$6 \times 10^{-25}~
{\rm W~m}^{-2}~{\rm s}^{-1}~{\rm sr}^{-1}$
for polarized emission
within each of 49152 equal-area pixels
and each of 400 spectral channels.
The resulting data set provides a rich resource for ancillary science.
PIXIE will measure distortions in the CMB blackbody spectrum
to test dark matter models
and extend the measurement of the scalar index $n_s$ of density perturbations
to scales as small as 1 kpc or 1 solar mass.
PIXIE measures both the temperature and ionization fraction
of the intergalactic medium at redshifts 5--30
to determine the nature of the first luminous objects in the universe.
PIXIE data in the far-infrared
determine the properties of the diffuse dust cirrus
and maps line emission from the molecules and ions that cool
the interstellar medium within the Galaxy.

\acknowledgments{
We thank J.~Feng for providing Figure \ref{slepton_fig}.
Figure \ref{line_ratio_fig} is reproduced by permission of the 
American Astronomical Society.
}

%------------------------------------------------------
% Appendix A: Optical Signal
%------------------------------------------------------
\appendix

\section{Optical Signal Path}
PIXIE is fully symmetric about the plane 
bisecting the transfer mirrors
(Figure \ref{pixie_concept_fig}).
Consider a plane wave incident on the instrument
A-side primary mirror,
\begin{equation}
\vec{E}_0 = E_x e^{i(kz - \omega t)} ~\hat{x}
  + E_y e^{i(kz - \omega t)} ~\hat{y}
\label{full_e0_eq}
\end{equation}
where
$\hat{x}$ is in the plane of the diagram,
$\hat{y}$ is normal to the page,
and
$\hat{z}$ is along the direction of propagation.
For clarity, we will ignore the term
$i(kz - \omega t)$
since it is common to all expressions,
and denote the amplitude of the field in the 
$\hat{x}$ and $\hat{y}$ directions as
\begin{equation}
\vec{E}_0 = A \hat{x} + B \hat{y}
\label{short_e0_eq}
\end{equation}
where $A = E_x$ and $B = E_y$.
Reflection from a mirror
reverses the direction of propagation
and flips the sign of the electric field.
Reflections from the primary, folding, and secondary mirrors
flip the sign three times and route the beam to the FTS.
Reflection from the first transfer mirror flips the sign a fourth time.
The field after reflection from the first transfer mirror is thus
\begin{equation}
\vec{E}_1 = \vec{E}_0 = A\hat{x} + B\hat{y}
\label{e1_eq}
\end{equation}
The first polarizing grid transmits the $\hat{x}$ polarization
while reflecting $\hat{y}$.
The second left-side transfer mirror
collects the beam reflected from the polarizing grid
(with a minus sign for the reflection),
while the right-side transfer mirror collects the transmitted beam.
After reflecting off the second transfer mirror
(which induces another sign flip),
the fields are
\begin{equation}
\vec{E}_{L2}= B\hat{y}~~~~~~~~~~~~~~~~~~ \vec{E}_{R2} = -A\hat{x}
\label{e2_eq}
\end{equation}
where subscripts $L$ and $R$ refer to the left and right sides
of the instrument, respectively.
The beams then encounter the second polarizing grid,
oriented such that from the point of view of the
radiation the wires are at 45\deg\ ~from the plane of polarization.
Half of each beam is transmitted and half is reflected,
imposing a new polarization basis on the radiation
which we denote
$\hat{u}$ and $\hat{v}$.
The relationship between $\hat{u},\hat{v}$ 
and $\hat{y},\hat{z}$ coordinate systems is
\begin{eqnarray}
\hat{u}=( \hat{x} + \hat{y}) / \sqrt{2}  \nonumber \\
\hat{v}=( \hat{x} - \hat{y}) / \sqrt{2}
\label{xy_uv_eq}
\end{eqnarray}
Hence the radiation just before the second polarizer
can be described as
\begin{eqnarray}
\vec{E}_{L2} = ~~B(\hat{u} - \hat{v}) / \sqrt{2}   \nonumber\\
\vec{E}_{R2} =  -A(\hat{u} + \hat{v}) / \sqrt{2}
\label{e2_uv_eq}
\end{eqnarray}
The wires of the second polarizer are aligned with the $\hat{u}$ vector
so that the E fields in the $\hat{u}$ direction are reflected 
(with a minus sign) 
while those in the $\hat{v}$ direction are transmitted. 

The third transfer mirror mirror 
on the left side
thus collects the reflected ($\hat{u}$) radiation from $\vec{E}_{L2}$
(with a minus sign)
plus the transmitted ($\hat{v}$) radiation from $\vec{E}_{R2}$.
The third transfer mirror on the right
collects transmitted ($\hat{v}$) radiation from $\vec{E}_{L3}$
plus the reflected ($\hat{u}$) radiation from $\vec{E}_{R3}$
(with a minus sign).
After reflection from the third transfer mirror
(which induces another sign flip),
the fields become
\begin{equation}
\vec{E}_{L3} = ( B\hat{u} + A\hat{v}) / \sqrt{2} ~~~~~~~~~~~
\vec{E}_{R3} = (-A\hat{u} + B\hat{v}) / \sqrt{2} .
\label{e3_uv_eq}
\end{equation}
In the original coordinate system we may write this as
\begin{equation}
\vec{E}_{L3} = [ (A+B) \hat{x} - (A-B) \hat{y}] / 2~~~~~~
\vec{E}_{R3} = [-(A-B) \hat{x} - (A+B) \hat{y}] / 2 .
\label{e3_xy_eq}
\end{equation}
Note that the second polarizing grid 
mixes the original polarization states --
we began with amplitude $A$ oriented along $\hat{x}$
and amplitude $B$ oriented along $\hat{y}$.
From Eq. \ref{e3_xy_eq}
we now have linear combinations of $A$ and $B$ 
along each coordinate axis $\hat{x}$ and $\hat{y}$.

Each beam then reflects from the dihedral mirror assembly.
The dihedral mirror treats the two polarization states differently.
The $\hat{x}$ polarization reflects from two faces,
generating two canceling negative signs, 
while the $\hat{y}$ polarization 
reflects from two faces but also changes direction,
resulting in a net change of sign.
In addition, 
the different path length of the left beam with respect to the right beam
generates an optical phase delay
that depends on the position of the dihedral mirror assembly.
The path is shortened on one side but lengthened on the other,
creating a total phase delay
$e^{+2i z \omega/c}$
on one side and
$e^{-2i z \omega/c}$
on the other side,
where $z$ is the position of the dihedral mirror assembly.
The beams then reflect from the fourth transfer mirror.
Including the phase delay, 
the electric fields after the fourth transfer mirror become
\begin{eqnarray}
\vec{E}_{L4} &=& [ -(A+B) \hat{x} - (A-B) \hat{y}]e^{2iz\omega /c}/2   \nonumber \\
\vec{E}_{R4} &=& [~~(A-B) \hat{x} - (A+B) \hat{y}]e^{-2iz\omega /c}/2 ,
\label{e4_eq_xy}
\end{eqnarray}
or in the rotated coordinate system,
\begin{eqnarray}
\vec{E}_{L4} &=& (-A \hat{u} - B \hat{v}) e^{ 2iz\omega /c}/\sqrt{2}   \nonumber \\
\vec{E}_{R4} &=& (-B \hat{u} + A \hat{v}) e^{-2iz\omega /c}/\sqrt{2} .
\label{e4_eq_uv}
\end{eqnarray}

\noindent
The beams then recombine at the third polarizing grid.
The phases are different (in general)
and will lead to either constructive or destructive interference.
The wires in the third polarizer are oriented identically
to the second polarizer.
Radiation leaving the fourth transfer mirror on the left side ($\vec{E}_{L4}$)
will thus reflect the $\hat{u}$ component 
to the fifth transfer mirror on the left side
(with a minus sign)
while transmitting the $\hat{v}$ component to the 
fifth transfer mirror on the right side.
The fifth transfer mirror induces another sign change.
The fields after the fifth transfer mirror are thus
\begin{eqnarray}
\vec{E}_{L5} &=& ( -A\hat{u}e^{2iz\omega /c} - A\hat{v}e^{-2iz\omega /c})
/\sqrt{2}\nonumber \\
\vec{E}_{R5} &=& ( -B\hat{u}e^{-2iz\omega/c} + B\hat{v}e^{2iz\omega/c})/\sqrt{2} ,
\label{e5_uv_eq}
\end{eqnarray}
or in the original coordinate system 
\begin{eqnarray}
\vec{E}_{L5} &=& [ -A (e^{2iz\omega /c} + e^{-2iz\omega/c}) ~\hat{x}
		   -A (e^{2iz\omega /c} - e^{-2iz\omega /c}) ~\hat{y} ~]/2  
		         \nonumber \\
\vec{E}_{R5} &=& [ ~~B (e^{2iz\omega /c} - e^{-2iz\omega/c}) ~\hat{x}
		    -B (e^{2iz\omega /c} + e^{-2iz\omega /c}) ~\hat{y} ~]/2  
\label{e5_xy_eq}
\end{eqnarray}
Re-writing the exponentials as sines and cosines,
we obtain
\begin{eqnarray}
\vec{E}_{L5} &=&     -A \cos(2z\omega /c) ~\hat{x}
	           -i A \sin(2z\omega /c) ~\hat{y} ~\nonumber \\
\vec{E}_{R5} &=&    i B \cos(2z\omega /c) ~\hat{x}
		    - B \cos(2z\omega /c) ~\hat{y}
\label{e5_eq_cos}
\end{eqnarray}
The beams then encounter the fourth polarizing grid.
The wires of the fourth grid are oriented the same as the first grid,
reflecting the $\hat{y}$ component while transmitting $\hat{x}$.
After a final reflection from the sixth transfer mirror,
the beams reach the detectors.
The electric fields at the detectors 
from radiation originally incident from the left-side beam 
are thus
\begin{eqnarray}
\vec{E}_{L6} & = &  -i B \sin(2z\omega /c) ~\hat{x}
		    -i A \sin(2z\omega /c) ~\hat{y}  \nonumber \\
\vec{E}_{R6} & = &   ~~A \cos(2z\omega /c) ~\hat{x}
		      -B \cos(2z\omega /c) ~\hat{y}
\label{e6_eq}
\end{eqnarray}

\noindent
PIXIE has 4 detectors,
each of which measures a single linear polarization.
The detectors are arranged in two pairs.
One member of each pair measures the $\hat{x}$ component
while the other member measures $\hat{y}$.
The bolometers measure the incident power,
which is the square of the electric field:
\begin{eqnarray}
P_{Lx} &=& B^2\sin^2(2z\omega /c) ~=~ \frac{1}{2} ~B^2[1-\cos(4z\omega /c)]	\nonumber \\
P_{Ly} &=& A^2\sin^2(2z\omega /c) ~=~ \frac{1}{2} ~A^2[1-\cos(4z\omega /c)]	\nonumber \\	
P_{Rx} &=& A^2\cos^2(2z\omega /c) ~=~ \frac{1}{2} ~A^2[1+\cos(4z\omega /c)]	\nonumber \\
P_{Ry} &=& B^2\cos^2(2z\omega /c) ~=~ \frac{1}{2} ~B^2[1+\cos(4z\omega /c)]
\label{left_p_eq}
\end{eqnarray}
The derivation thus far has ignored radiation incident from the B-side beam.
Since PIXIE is fully symmetric,
the derivation for radiation from the B-side beam 
is identical to radiation from the A-side beam.
Using the principle of superposition
we can combine the results from both sides and all frequencies
to obtain the power incident on the bolometers
as a function of the dihedral mirror position $z$:
\begin{eqnarray}
P_{Lx} &=& \frac{1}{2} ~\int(B^2+C^2)+(C^2-B^2) \cos(4z\omega /c)d\omega    \nonumber \\
P_{Ly} &=& \frac{1}{2} ~\int(A^2+D^2)+(D^2-A^2) \cos(4z\omega /c)d\omega    \nonumber \\
P_{Rx} &=& \frac{1}{2} ~\int(A^2+D^2)+(A^2-D^2) \cos(4z\omega /c)d\omega    \nonumber \\
P_{Ry} &=& \frac{1}{2} ~\int(B^2+C^2)+(B^2-C^2) \cos(4z\omega /c)d\omega    
\label{full_p_eq_2}
\end{eqnarray}
where we use notation
\begin{eqnarray}
A &=& E_x^A ~=~ \hat{x}~{\rm component~of~incident~E~field~from~A-side~beam}    \nonumber \\
B &=& E_y^A ~=~ \hat{y}~{\rm component~of~incident~E~field~from~A-side~beam}    \nonumber \\
C &=& E_x^B ~=~ \hat{x}~{\rm component~of~incident~E~field~from~B-side~beam}   \nonumber \\
D &=& E_y^B ~=~ \hat{y}~{\rm component~of~incident~E~field~from~B-side~beam} . \nonumber
\label{abcd_def}
\end{eqnarray}

\noindent
Each detector thus measures a DC term
plus an interference fringe pattern
modulated by the position of the dihedral mirror.
Low-frequency noise will render the DC component unusable.
The modulated term is proportional to the Fourier transform
of the difference spectrum
between one linear polarization from one input beam
and the orthogonal linear polarization from the other beam.

PIXIE can operate in two modes.
With the calibrator stowed, 
both beams view the same part of the sky.
The power on each detector then becomes
\begin{eqnarray}
P_{Lx} = \frac{1}{2} ~\int 
  ( E_x^2 + E_y^2 )
+ ( E_x^2 - E_y^2 ) \cos(4z\omega /c) ~d\omega    	\nonumber \\
P_{Ly} = \frac{1}{2} ~\int 
  ( E_x^2 + E_y^2 )
- ( E_x^2 - E_y^2 ) \cos(4z\omega /c) ~d\omega    	\nonumber \\
P_{Rx} = \frac{1}{2} ~\int 
  ( E_x^2 + E_y^2 )
+ ( E_x^2 - E_y^2 ) \cos(4z\omega /c) ~d\omega    	\nonumber \\
P_{Ry} = \frac{1}{2} ~\int
  ( E_x^2 + E_y^2 )
- ( E_x^2 - E_y^2 ) \cos(4z\omega /c) ~d\omega .
\label{det_q_mode}
\end{eqnarray}
We may rewrite this in terms of the Stokes parameters
\begin{eqnarray}
I &=& E_x^2 + E_y^2 	\nonumber \\
Q &=& E_x^2 - E_y^2	\nonumber \\
U &=& 2 E_x E_y	
\label{stokes_def}
\end{eqnarray}
to obtain
\begin{eqnarray}
P_{Lx} = \frac{1}{2} ~\int ~I(\omega) + Q(\omega) \cos(4z\omega /c) ~d\omega    	\nonumber \\
P_{Ly} = \frac{1}{2} ~\int ~I(\omega) - Q(\omega) \cos(4z\omega /c) ~d\omega    	\nonumber \\
P_{Rx} = \frac{1}{2} ~\int ~I(\omega) + Q(\omega) \cos(4z\omega /c) ~d\omega    	\nonumber \\
P_{Ry} = \frac{1}{2} ~\int ~I(\omega) - Q(\omega) \cos(4z\omega /c) ~d\omega 
\label{det_q_mode_iq}
\end{eqnarray}
The detected fringe pattern
is thus directly proportional to the
Fourier transform of the 
frequency spectrum of the 
Stokes Q linear polarization of the sky.
The spacecraft spin
modulates the polarized signal from the sky
while leaving the unpolarized component unchanged.
The sky signal
\begin{equation}
\vec{E} = E_x \hat{x} + E_y \hat{y}
\label{e_sky_eq}
\end{equation}
in sky-fixed coordinates 
$[\hat{x}, \hat{y}]$
becomes
\begin{eqnarray}
\vec{E} &=& ( E_x \cos\gamma + E_y \sin\gamma ) ~\hat{x}^\prime \nonumber \\
        &+& ( E_y \cos\gamma - E_x \sin\gamma ) ~\hat{y}^\prime
\label{e_inst_eq_2}
\end{eqnarray}
in instrument coordinates
$[\hat{x}^\prime, \hat{y}^\prime]$,
where $\gamma$ is the spacecraft rotation angle
relating the two coordinate systems.
When the calibrator is in the stowed position,
we may thus recover the polarization state of the sky as
\begin{eqnarray}
P_{Lx} &=& \frac{1}{2} ~\int 
 [  Q(\omega) \cos(2\gamma) + U(\omega) \sin(2\gamma) ] \cos(4z\omega /c) ~d\omega    	\nonumber \\
P_{Ly} &=& \frac{1}{2} ~\int 
 [ -Q(\omega) \cos(2\gamma) - U(\omega) \sin(2\gamma) ] \cos(4z\omega /c) ~d\omega    	\nonumber \\
P_{Rx} &=& \frac{1}{2} ~\int 
 [ Q(\omega) \cos(2\gamma) + U(\omega) \sin(2\gamma) ] \cos(4z\omega /c) ~d\omega    	\nonumber \\
P_{Ry} &=& \frac{1}{2} ~\int 
 [ -Q(\omega) \cos(2\gamma) - U(\omega) \sin(2\gamma) ] \cos(4z\omega /c) ~d\omega,
\label{det_q_mode_iqu}
\end{eqnarray}
where we have dropped for clarity the constant term 
(proportional to Stokes I)
not modulated by the mirror movement.

The instrument can also operate 
with a blackbody calibrator blocking one aperture.
Assume that the calibrator blocks the A-side beam.
The detected power may now be written
\begin{eqnarray}
P_{Lx} &=& \frac{1}{2} ~\int 
  ( E_{x,{\rm sky}}^2 + E_{y,{\rm cal}}^2 )
+ ( E_{x,{\rm sky}}^2 - E_{y,{\rm cal}}^2 ) \cos(4z\omega /c) ~d\omega    	\nonumber \\
P_{Ly} &=& \frac{1}{2} ~\int 
  ( E_{x,{\rm cal}}^2 + E_{y,{\rm sky}}^2 )
- ( E_{x,{\rm cal}}^2 - E_{y,{\rm sky}}^2 ) \cos(4z\omega /c) ~d\omega    	\nonumber \\
P_{Rx} &=& \frac{1}{2} ~\int 
  ( E_{x,{\rm cal}}^2 + E_{y,{\rm sky}}^2 )
+ ( E_{x,{\rm cal}}^2 - E_{y,{\rm sky}}^2 ) \cos(4z\omega /c) ~d\omega    	\nonumber \\
P_{Ry} &=& \frac{1}{2} ~\int
  ( E_{x,{\rm sky}}^2 + E_{y,{\rm cal}}^2 )
- ( E_{x,{\rm sky}}^2 - E_{y,{\rm cal}}^2 ) \cos(4z\omega /c) ~d\omega    	\nonumber \\
\label{det_i_mode}
\end{eqnarray}
The fringe pattern now depends on the difference
between the sky signal in one linear polarization
and the calibrator signal in the orthogonal polarization.
We may write a single linear polarization state
as a linear combination of Stokes parameters,
\begin{eqnarray}
E_{x,{\rm sky}}^2 &=& \frac{ I_{\rm sky} + Q_{\rm sky} }{2}	\nonumber \\
E_{y,{\rm sky}}^2 &=& \frac{ I_{\rm sky} - Q_{\rm sky} }{2}	\nonumber \\
\label{pol_def}
\end{eqnarray}
The calibrator is unpolarized so that 
$\langle E_{x,{\rm cal}}^2 \rangle = \langle E_{y,{\rm cal}}^2 \rangle$.
As the spacecraft spins,
the measured fringe pattern
yields the frequency spectrum of the
Stokes I, Q, and U parameters from the sky,
\begin{eqnarray}
P_{Lx} &=& \frac{1}{4} ~\int \left[~
   ~I_{\rm sky}(\omega) - I_{\rm cal}(\omega)
+  Q(\omega) \cos(2\gamma) + U(\omega) \sin(2\gamma) 
~\right] \cos(4z\omega /c) ~d\omega 	\nonumber \\
P_{Ly} &=& \frac{1}{4} ~\int \left[~
   ~I_{\rm sky}(\omega) - I_{\rm cal}(\omega)
-  Q(\omega) \cos(2\gamma) - U(\omega) \sin(2\gamma) 
~\right] \cos(4z\omega /c) ~d\omega 	\nonumber \\
P_{Rx} &=& \frac{1}{4} ~\int \left[~
   ~I_{\rm cal}(\omega) - I_{\rm sky}(\omega)
+  Q(\omega) \cos(2\gamma) + U(\omega) \sin(2\gamma) 
~\right] \cos(4z\omega /c) ~d\omega 	\nonumber \\
P_{Ry} &=& \frac{1}{4} ~\int \left[~
   ~I_{\rm cal}(\omega) - I_{\rm sky}(\omega)
-  Q(\omega) \cos(2\gamma) - U(\omega) \sin(2\gamma) 
~\right] \cos(4z\omega /c) ~d\omega ,
\label{det_i_mode_iqu}
\end{eqnarray}
where we again drop for clarity the constant terms
(now proportional to $I_{\rm cal} + I_{\rm sky} + Q_{\rm sky}$)
not modulated by the mirror movement.
Note the factor of two difference for the Stokes Q and U terms
between Eqs. \ref{det_q_mode_iqu} and \ref{det_i_mode_iqu}.
When the calibrator is over one beam,
the fringe pattern is sensitive to Stokes I, Q, and U.
When both beams view the sky,
the fringe pattern is only sensitive to linear polarization (Stokes Q and U),
but at twice the signal intensity
since the instrument now interferes two copies of the sky signal.

The factor of two increase in sensitivity 
when both beams view the sky
is possible because the instrument
interferes one linear polarization from one beam
with the orthogonal polarization from the other beam.
Circularly polarized sky emission
would introduce correlations 
between the orthogonal linear polarization states,
but continuum emission at millimeter wavelengths
has no significant circular polarization.

%------------------------------------------------------
% Appendix B: Noise Budget
%------------------------------------------------------

\section{Noise Budget}
Several sources of noise contribute to the total noise budget.
The dominant term is statistical noise from the incident photons.
The instrument is contained in an emissive cavity
maintained at the same temperature as the CMB.
For noise from a CMB source
we set the emissivity $\epsilon=1$ and throughput $f=1$
in Eq. \ref{mather_12a},
since any CMB photon absorbed by the instrument
will be replaced by an equivalent photon
emitted by the instrument.
The etendu $A \Omega$ in Table \ref{optic_param}
refers to the circularized beam
truncated by the field and pupil stops.
Since each detector views these stops,
their emission also contributes to the photon noise.
We conservatively assume that the beam stops
are at the same 2.725 K temperature as the rest of the optics
and increase the effective etendu by a factor 1.27
in Eq. \ref{nep_cmb_val}
to account for emission from the beam stops.
We numerically evaluate Eq. \ref{mather_12a},
yielding CMB photon noise
\begin{equation}
{\rm NEP}_{\rm CMB} = 2.0 \times 10^{-16} ~{\rm W~Hz}^{-1/2} .
\label{nep_cmb_val}
\end{equation}

At sub-millimeter wavelengths,
photon noise from the Galactic dust cirrus is non-negligible.
We model the mid-latitude dust
as a modified greybody,
\begin{equation}
I(\nu)_{\rm dust} = \epsilon B_\nu(T_{\rm dust}) 
	\left( \frac{\nu}{\nu_0} \right)^\beta ,
\label{dust_intensity}
\end{equation}
with values
$\epsilon = 2 \times 10^{-4}$,
$T_{\rm dust} = 18$ K,
$\beta = 2$,
and reference frequency $\nu_0 = 3$ THz
\citep{finkbeiner/etal:1999},
where $B_\nu(T)$ is the Planck function.
The instrument cavity does not contribute to the THz photon noise.
We thus use optical parameters from Table \ref{optic_param}
to obtain
\begin{equation}
{\rm NEP}_{\rm dust} = N ~1.6 \times 10^{-16} ~{\rm W~Hz}^{-1/2},
\label{nep_dust_val}
\end{equation}
where the constant
$N=1$ when both apertures are open to the sky
and
$N=1/2$ when the calibrator blocks one aperture
({\it cf} the factors $1/2$ and $1/4$ in 
Eqs. \ref{det_q_mode_iq} and \ref{det_i_mode_iqu}).

A third source of noise 
is the intrinsic phonon noise associated with the detector,
measured as
\begin{equation}
{\rm NEP}_{\rm phonon} = 0.7 \times 10^{-16} ~{\rm W~Hz}^{-1/2} .
\label{nep_phonon}
\end{equation}
PIXIE will thus operate with nearly photon-limited noise
given by the quadrature sum of the photon and phonon terms,
\begin{eqnarray}
{\rm NEP} &=& 2.3 \times 10^{-16} ~{\rm W ~Hz}^{-1/2} ~~~~~{\rm (Calibrator~deployed)} \nonumber \\
          &=& 2.7 \times 10^{-16} ~{\rm W ~Hz}^{-1/2} ~~~~~{\rm (Calibrator ~stowed)} .
\label{asp_nep}
\end{eqnarray}
The larger dust signal
when the calibrator is stowed
increases the photon noise in this configuration.
However, the small increase in photon noise is compensated
by a doubling of the sky signal,
so that the overall system sensitivity improves
when the calibrator is stowed
(Eqs. \ref{spectral_noise_i_mode} and \ref{spectral_noise_q_mode}).

Photon noise from the CMB is isotropic on the sky,
but the contribution from dust will vary with position.
Equation \ref{nep_dust_val} 
uses the mean dust amplitude for Galactic latitude $|b| > 30\deg$.
The sensitivity will be slightly better
in regions of low dust signal
and slightly worse near the Galactic plane.
Note that the sharp edges of the PIXIE tophat beam
minimize spillover of Galactic plane emission
to higher latitudes.

The detector noise and sampling determine the
noise in the time-ordered data.
The data taken as the mirror moves 
from one end of its throw to the other end
constitutes a single interferogram,
which we Fourier transform
to derive the difference spectrum
between the two beams.
We assume each interferogram uses
$N_s$ samples
and define the forward Fourier transform as
\begin{equation}
S_\nu = \sum_{k=0}^{N_s-1}{ S_i \exp(2\pi i \nu k / N_s )}
\label{fft_def}
\end{equation}
where
$S_i$ is a time-ordered sample
and $\nu$ denotes frequency.
The noise in each time-ordered sample is
\begin{equation}
\delta S_i = \sqrt{ \frac{2}{\delta t} } ~{\rm NEP} 
\label{dp_tod_eq}
\end{equation}
where
$\delta t ~\sim ~1$ ms
is the integration time per sample
and the factor of 2 accounts
for the Nyquist sampling
between the frequency and time domains.
The noise from different time-ordered bins is uncorrelated
and will add incoherently in the FFT,
while the signal is correlated 
and adds coherently.
The RMS noise in a single bin of the Fourier transform
is thus
\begin{equation}
\delta S_\nu = \frac{\delta S_i}{\sqrt{N_\nu}}
\label{dp_f_eq}
\end{equation}
where $N_\nu = N_s/2$ is the number of discrete frequencies 
in the FFT.
Assume we observe the sky
for a time $t_{\rm IFG}$
corresponding to a single interferogram
(physical movement of the mirror 
from one endpoint through the null to the other endpoint).
If there are $N_s$ samples per interferogram
then each sample is observed for a total integration time
$\delta t = t_{\rm IFG} / N_s$.
The noise in each bin of the FFT derived from a single interferogram 
for each detector
then becomes
\begin{equation}
\delta S_{\nu}^{\rm IFG} = \frac{2 ~{\rm NEP}}{ \sqrt{ t_{\rm IFG} }}
\label{dp_pix_eq}
\end{equation}
independent of the number of frequency bins.
This result is independent of the
choice of normalization in the Fourier transform.
If we instead adopted the reverse transform
$S_\nu = \frac{1}{N_s} \sum_{k=0}^{N_s-1}{ S_i \exp(2\pi i \nu k / N_s )}$,
we would obtain an additional compensating factor of $N_s$
when deriving the sky signal in Eq. \ref{i_noise}.

% -------------- Table 5: Dihedral Apodization --------------
\begin{table}[t]
{
\small
\caption{Dihedral Apodization Sampling}
\label{scan_table}
\begin{center}
\begin{tabular}{| c | c | c | c | c |}
\hline 
Orbit 	& Optical & Physical & Samples     & Strokes  \\
Number  & Delay   & Stroke   & per Stroke  & per Spin \\	
\hline 
1 \& 2	& $\pm$10 mm	& $\pm$2.6 mm	& 1024  & 8   \\
3 \& 4	& $\pm$8.9 mm	& $\pm$2.3 mm	& 910	& 9   \\
5 \& 6	& $\pm$8.0 mm	& $\pm$2.1 mm	& 819	& 10  \\
7 \& 8	& $\pm$6.7 mm	& $\pm$1.7 mm	& 683	& 12  \\
9 \& 10	& $\pm$5.0 mm	& $\pm$1.3 mm	& 512	& 16  \\
11 \& 12& $\pm$3.3 mm	& $\pm$0.9 mm	& 341	& 24  \\
\hline
\end{tabular}
\end{center}
}
\end{table}
%------------------------------------------------------------

The sky signal is split among the 4 detectors,
effectively increasing the noise by a factor of 4.
When the calibrator is stowed,
the instrument interferes two copies of the sky signal,
gaining back a factor of 2
({\it cf} the factors $1/4$ and $1/2$
in Eqs. \ref{det_q_mode_iqu} and \ref{det_i_mode_iqu}).
In addition, the raw interferograms
are a linear combination
of polarized and unpolarized sky signal.
We Fourier transform the interferogram from a single detector
and bin the resulting spectrum by spacecraft rotation angle $\gamma$,
to fit for the Stokes $I$, $Q$, and $U$ parameters in each sky pixel.
The $\sin(2\gamma)$ or $\cos(2\gamma)$ basis functions
increase the noise for the fitted
Q or U parameters by a factor $\sqrt 2$.

If the instrument throughput were independent of frequency,
the noise in the specific intensity spectra 
would also be independent of frequency.
Several factors reduce the throughput at high frequencies.
The beams within the FTS are not plane waves
but use collimating optics with half-angle 6\ddeg5.
The resulting dispersion causes loss of fringe coherence
at short wavelengths.
The optical polarizers use wire grids,
which become inefficient in reflection
at wavelengths shorter than twice the 30 $\mu$m wire pitch,
resulting in a loss of fringe amplitude.
We model this as a single-pole filter
with knee at twice the grid pitch.
There are 4 identical grids,
resulting in a 4-pole low-pass filter.
A separate low-pass filter 
with a knee at 100 $\mu$m
blocks short-wavelength zodiacal emission
to prevent radiation
at frequencies beyond the highest synthesized spectral channel
from affecting the photon noise budget.
Finally, the doped silicon absorbing structure on the
polarization-sensitive bolometers
loses efficiency at short wavelengths.
We model this as an additional low-pass filter
with knee at 60 $\mu$m.
All sensitivity estimates include these low-pass filters.
The primary effect is to limit the effective spectral range
of the instrument.
Although the synthesized spectra have 512 channels
to a maximum of 7665 GHz,
channels at frequencies above 6 THz
(wavelength 50 $\mu$m) will have little sensitivity to sky signal.

%--------------------------------------------------------------------------
% Figure 17: MTM apodization cartoon
%--------------------------------------------------------------------------
\begin{figure}[t]
\centerline{
\includegraphics[height=3.0in]{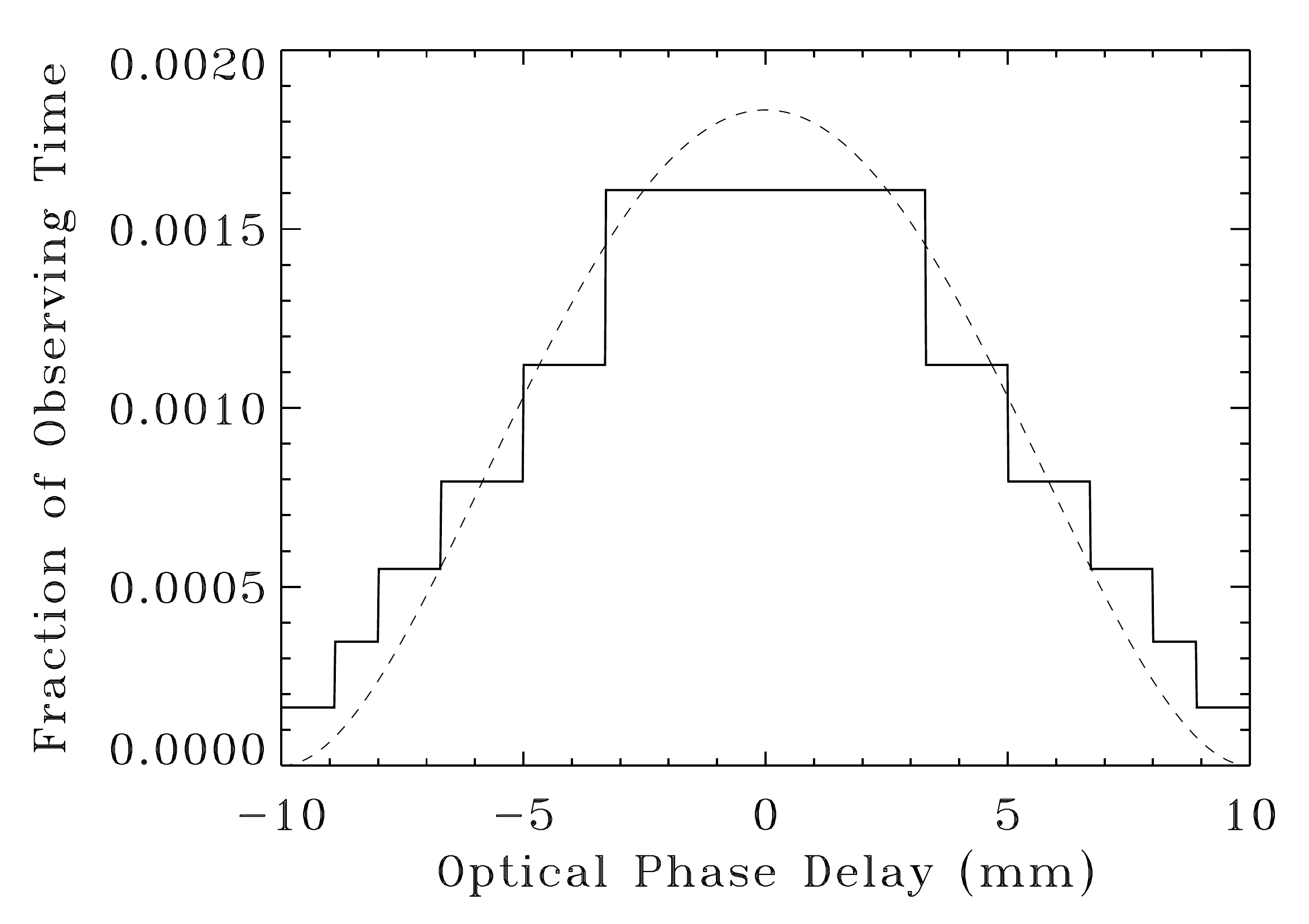}
}
\caption{
Apodization of the optical phase delay $\Delta L$
created by scanning the dihedral mirror through different physical stroke lengths
on different orbits.
The achieved apodization (solid line)
closely approximates the ideal (dashed line).}
\label{apodization_fig}
\end{figure}
%--------------------------------------------------------------------------

%------------------------------------------------------
% Appendix C: Fourier Apodization
%------------------------------------------------------

\section{Fourier Apodization}
A Fourier transform uniformly sampled from -1 to +1
has sharp edges at $\pm$1,
leading to ringing in the frequency domain.
In addition, the CMB only produces fringes near zero path length ---
observations at longer optical path,
necessary to provide sufficiently narrow frequency bins,
have little CMB signal and contribute little to the CMB sensitivity.
We minimize ringing and maximize sensitivity
by varying the mirror stroke to apodize the Fourier transform.
This softens the edges in the position domain
by spending less time observing at large optical path length,
while maximizing CMB sensitivity
by spending more time observing CMB fringes near the optical null.
A commonly used apodization is $(1 - z^2)^2$.
We approximate this apodization
by varying the mirror stroke length on successive orbits.
PIXIE observes the sky from a polar sun-synchronous orbit.
The orbit plane precesses 1\deg ~per day,
so that a pixel near the celestial equator
is visible for at least 13 consecutive orbits.
Each orbit uses a different mirror stroke length,
ranging from shortest stroke 
$\Delta z = \pm 0.86$ mm
(optical path length $\Delta L = \pm 3.3$ mm)
to a longest stroke
$\Delta z = \pm 2.58$ mm
(optical path length $\Delta L = \pm 10.0$ mm).
Table \ref{scan_table}
summarizes the mirror stroke pattern.
The resulting apodization closely approximates the ideal
(Figure \ref{apodization_fig}).
Optical paths near zero delay are observed much more often
than paths at large delay,
and have correspondingly smaller noise,
while optical paths at large phase delay
are seldom observed 
and have larger noise.
Simulations of the Fourier transform
using this non-uniform noise pattern
show noise in the Fourier transform
to be smaller by a factor 0.73
compared to simulations
with the equivalent integration time
and no apodization.

% -------------- References --------------
% To keep things simple, list the references below in the order in which
% they are cited in the text.
%-----------------------------------------

% That's all there is, kiddies.  Ride off into the sunset!
\end{document}